\documentclass[12pt]{article}
\usepackage{cite}
\usepackage[dvips]{graphicx}
\usepackage{epsfig}
\usepackage{amssymb}
\usepackage{rotating}
\usepackage{mathtools}

\def \Rpv{R_{P} \hspace{-1.2em}/\;\hspace{0.2em}}
\def \emass {\langle m_{\nu} \rangle}
\def \znbb {0\nu\beta\beta}

\def \Rpv{R_{P} \hspace{-1.2em}/\;\hspace{0.2em}}
\def\rpm{R_p \hspace{-0.8em}/\;\:}
\def\pslash{p \hspace{-0.5em}/\;\:}
\def\gsim{\raise0.3ex\hbox{$\;>$\kern-0.75em\raise-1.1ex\hbox{$\sim\;$}}}
\def\lsim{\raise0.3ex\hbox{$\;<$\kern-0.75em\raise-1.1ex\hbox{$\sim\;$}}}

\renewenvironment{thebibliography}[1]{%
\begin{oldthebibliography}{#1}%
\setlength{\parskip}{0ex}%
\setlength{\itemsep}{0ex}%
}%
{%
\end{oldthebibliography}%
}

\begin{document}

\vspace*{-1in}
\renewcommand{\thefootnote}{\fnsymbol{footnote}}

\begin{flushright}
%Draft April 1999\\
\texttt{%hep-ph/???
}
\end{flushright}
\vskip 5pt

\begin{center}
{\Large{\bf 
Neutrinoless Double Beta Decay and \\[0.2cm]
Physics Beyond the Standard Model}}
\vskip 25pt

{
Frank F. Deppisch$^1$\footnote{E-mail: f.deppisch@ucl.ac.uk},
Martin Hirsch$^2$\footnote{E-mail: mahirsch@ific.uv.es} and
Heinrich P\"as$^3$\footnote{E-mail: heinrich.paes@uni-dortmund.de}
}

\vskip 10pt

{\it \small $^1$Department of Physics and Astronomy, University College London, UK}\\
{\it \small $^2$AHEP Group, Instituto de F\'isica Corpuscular, Universitat de Valencia, Spain}\\
{\it \small $^3$Fakult\"at f\"ur Physik, Technische Universit\"at Dortmund, Germany}\\

\medskip

\begin{abstract}
\noindent
Neutrinoless double beta decay is the most powerful tool to probe not
only for Majorana neutrino masses but for lepton number violating
physics in general. We discuss relations between lepton number
violation, double beta decay and neutrino mass, review a general
Lorentz invariant parametrization of the double beta decay rate,
highlight a number of different new physics models showing how
different mechanisms can trigger double beta decay, and finally
discuss possibilities to discriminate and test these models
and mechanisms in complementary experiments.
\end{abstract}

\end{center}

\medskip

\noindent
{\small PACS numbers: 23.40.BW, 11.30.Fs 14.80 }

\renewcommand{\thefootnote}{\arabic{footnote}}
\setcounter{footnote}{0}

%------------------------------------------------------------------------
\section{Introduction}
\label{sec:Introduction}
%------------------------------------------------------------------------

The search for neutrinoless double beta decay ($0\nu\beta\beta$) - the simultaneous
transformation of two neutrons into two protons, two electrons and
nothing else - is the most sensitive tool for probing Majorana
neutrino masses (see the contribution by Rodejohann
\cite{Rodejohann:2012xd}). However, while this so-called mass
mechanism is certainly the most prominent realization of the decay,
and while an uncontroversial detection of neutrinoless double beta
decay will inevitably guarantee that neutrinos are Majorana particles,
Majorana neutrino masses are not the only element of beyond Standard
Model physics which can induce double beta.

In this review we discuss mechanisms of neutrinoless double beta
decay where the lepton number violation (LNV), necessary for the decay, does
not directly originate from Majorana neutrino masses but rather due to
lepton number violating masses or couplings of new particles appearing
in various possible extensions of the Standard Model. While the same
couplings will also induce Majorana neutrino masses, due to the
Schechter-Valle black box theorem \cite{Schechter:1981bd, Nieves:1984sn}, 
in these cases the double beta decay half life will not yield any {\em direct} information about the neutrino mass.

We start our review with a general consideration of black box
contributions to neutrinoless double beta decay and neutrino
masses. Next, we discuss a general framework which allows to
parametrize and analyze any single contribution to neutrinoless double
beta decay allowed by Lorentz invariance.  While the neutrino mass
limit is based on the well-known mechanism exchanging a massive
Majorana neutrino between two standard model ($V-A$) vertices, the
effective vertices appearing in the new contributions involve
non--standard currents such as scalar, pseudoscalar and tensor
currents. Moreover, besides contributions with a light neutrino being
exchanged between two separated vertices, the so-called long-range
part, additional contributions from short range mechanisms are possible,
where exchanged particles are all much heavier than the typical length
scale of nuclear separation, such as in supersymmetry (SUSY) without 
$R$--parity.
Then we turn to several concrete models such as left--right symmetry
\cite{Doi:1985dx, Hirsch:1996qw}, $R$--parity violating SUSY
\cite{Hirsch:1995zi, Hirsch:1995ek, Babu:1995vh, Hirsch:1995cg, Pas:1998nn} and leptoquarks \cite{Hirsch:1996qy, Hirsch:1996ye}. Finally we discuss the prospects to
discriminate different mechanisms.

%------------------------------------------------------------------------
\section{Black Box Theorem}
\label{sec:blackboxtheorem}
%------------------------------------------------------------------------

\begin{figure}
\centering
\includegraphics[clip,width=0.4\textwidth]{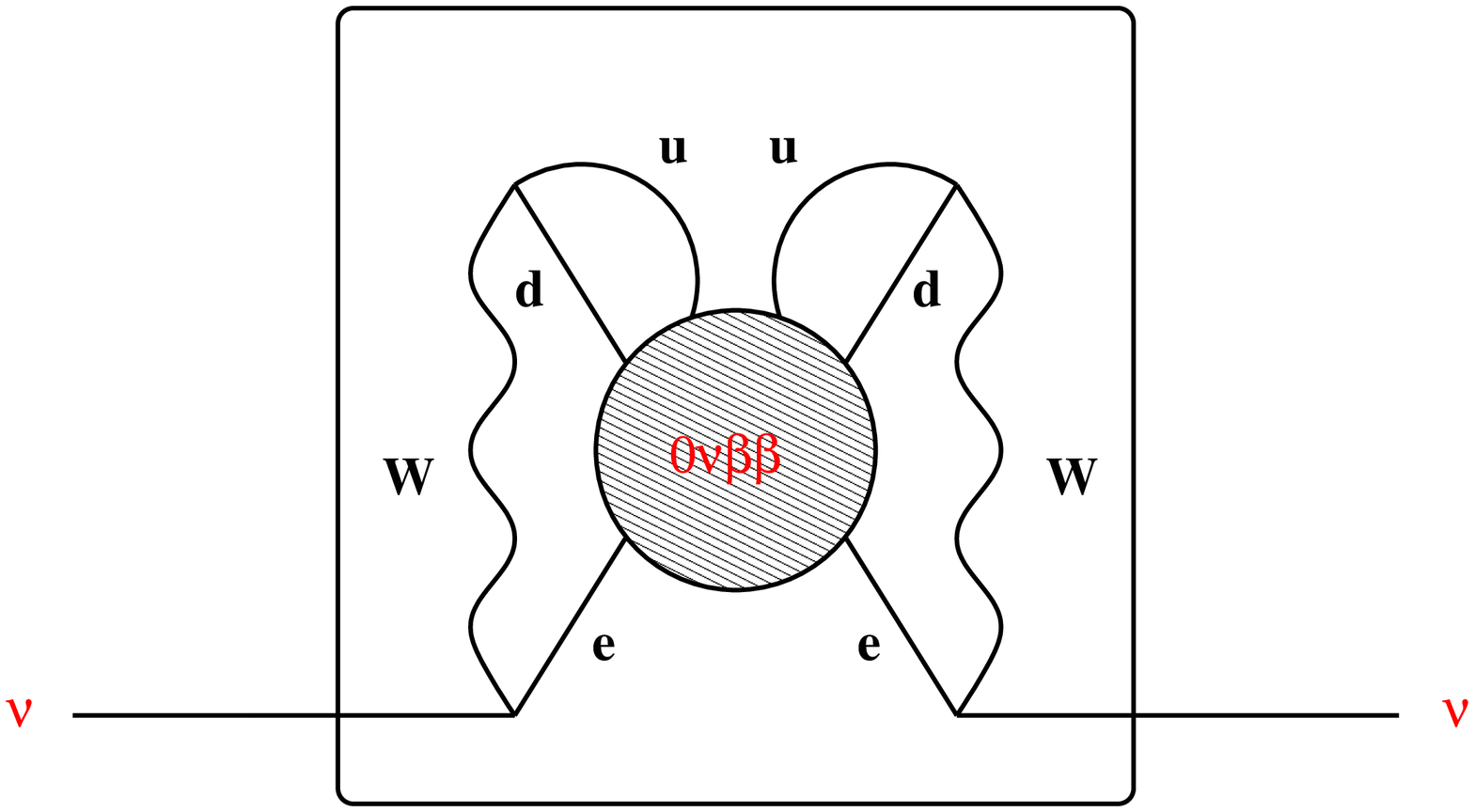}
\includegraphics[clip,width=0.4\textwidth]{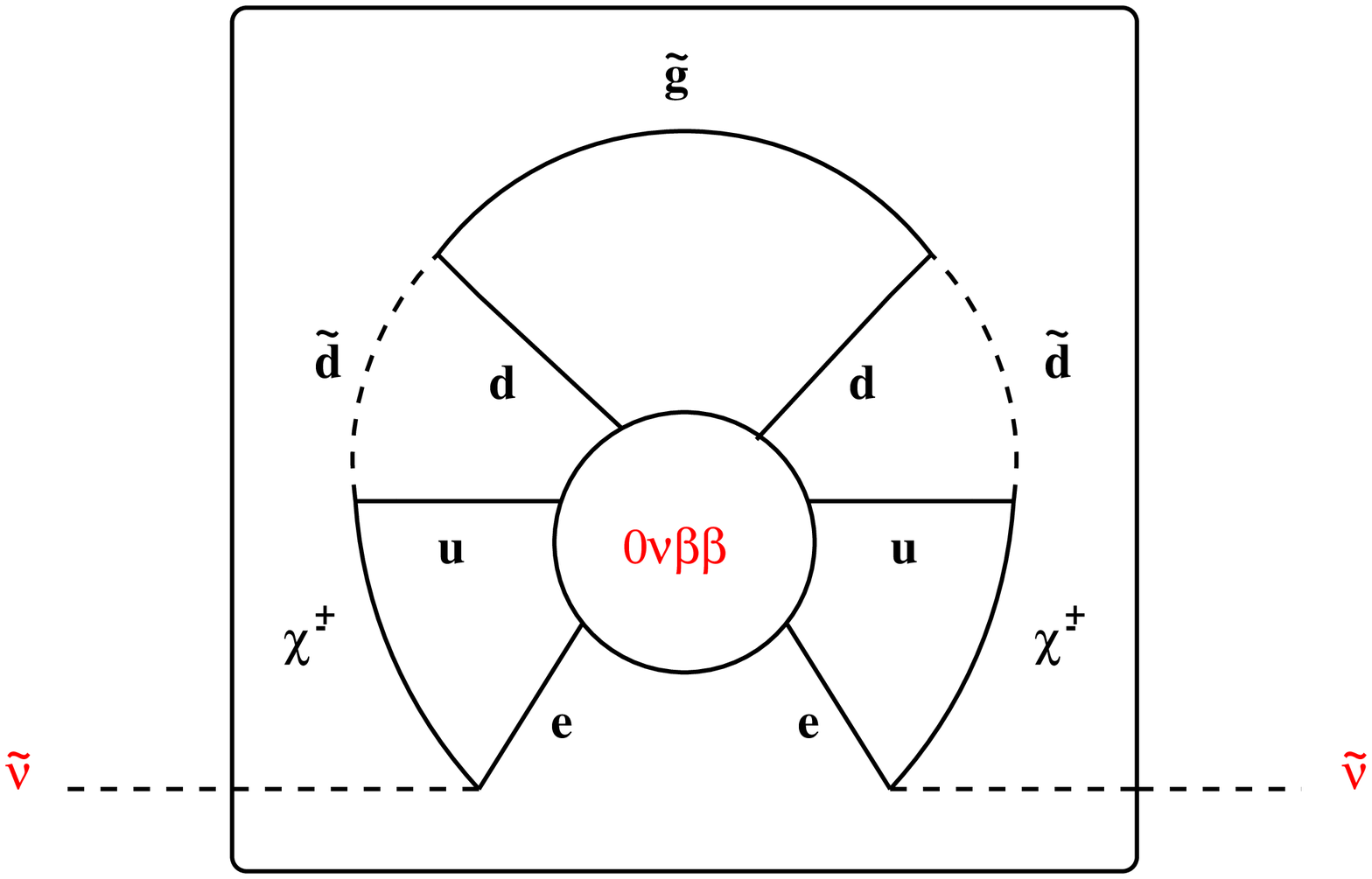}
\caption{\label{fig:bbth}Black box theorem graphically. To the 
left the original non-supersymmetric ``butterfly'' diagram, to 
the right the supersymmetric theorem.}
\end{figure}

The observation of neutrinoless double beta decay demonstrates that
lepton number is violated. Lepton number violation implies that
neutrinos have to be Majorana particles. That the two are inseparably
connected can be proven by what is known as the black box theorem
\cite{Schechter:1981bd, Nieves:1984sn, Takasugi:1984xr}. Graphically
the theorem can be depicted as shown in Fig.~\ref{fig:bbth}~(left): If double beta decay has been seen, a Majorana neutrino mass
term is generated at higher loop order, even if the underlying
particle physics model does not contain a tree-level neutrino mass.

One might wonder, if it is possible to circumvent the connection
between Majorana neutrino mass and double beta decay by tuning, for
example, in a given model the tree-level and 1-loop contributions to
the neutrino mass in such a way that the resulting {\em observable}
neutrino mass is too small to be detected and this is of course
possible. However, a tuning order-by-order in perturbation theory
requires a stabilizing symmetry for it to be natural and the black box
theorem \cite{Schechter:1981bd} states that no such symmetry can
exist, as has been shown formally in \cite{Takasugi:1984xr}.

A word of caution. The black box theorem has often been mis-presented 
in the literature. It does not guarantee by any means that the 
mass mechanism of neutrinoless double beta decay is dominant. This 
can be seen already by looking at the ``butterfly'' diagram in 
Fig.~\ref{fig:bbth}~(left): If this diagram is the only contribution to 
the neutrino mass the resulting $m_{\nu}$ is nearly infinitesimal, 
due to the 4-loop suppression factor. A recent calculation of the 
diagram in \cite{Duerr:2011zd} quotes a neutrino mass of the order 
of $m_{\nu} \sim 10^{-24}$ eV from current double beta decay limits. 
However, this does not imply that the neutrino mass has always to be 
this tiny in order for non-mass-mechanism contributions to double 
beta decay to dominate. In fact it is easy to find examples for both 
kinds of particles physics models, those that do give mass mechanism 
dominance and those that do not. The classic example of the former 
is the seesaw mechanism, which leaves (in a non-supersymmetric world) 
double beta decay dominated by neutrino mass {\em as the only experimental 
signature}. In the opposite class falls supersymmetry with (trilinear) 
$R$--parity violation, see below. 

One can also prove a supersymmetric version of the black box theorem
\cite{Hirsch:1997vz,Hirsch:1997dm,Hirsch:1997is}. This is depicted
graphically in Fig.~\ref{fig:bbth}~(right). In a supersymmetric
theory the scalar partner of the ordinary neutrino is a complex
field. Once lepton number is violated, this complex field splits into
its real and imaginary components with a non-zero mass difference
${\tilde m}_M^2$. This mass splitting generates lepton number violating effects such as sneutrino to anti-sneutrino oscillations
\cite{Grossman:1997is, Hirsch:1998gr}.  In analogy with the ordinary
black box theorem one can show that such a LNV mass splitting in the
sneutrino sector leads to double beta decay and observation of double
beta decay implies that ${\tilde m}_M^2$ is different from zero. At
the same time, Majorana neutrino masses and ${\tilde m}_M^2$ are also
always connected and, as shown in
\cite{Hirsch:1997vz,Hirsch:1997dm,Hirsch:1997is}, the existence of one
implies the existence of the other (if SUSY indeed is realized).

Finally, we mention that the original version of the black box theorem
\cite{Schechter:1981bd, Nieves:1984sn, Takasugi:1984xr} considers only
the first generation of leptons.  Oscillation experiments have shown,
however, that flavour violation exists in the neutrino sector. It is
then possible to prove an extended black box theorem
\cite{Hirsch:2006yk}, which takes also into account flavour. In
essence, this theorem states that under the assumption that all three
light neutrinos are Majorana particles, current oscillation data
requires the $\znbb$ decay observable $m_{ee}$, see
eq. (\ref{eq:defmeff}) below, must be different from zero, since no
symmetry can exist which guarantees this entry in the neutrino mass
matrix to vanish exactly. While academically amusing, however, the
theorem can not predict the actual value of $m_{ee}$, exactly as the
original version can not guarantee the dominance of the mass
mechanism. Thus, the results of \cite{Hirsch:2006yk} do not guarantee
that the double beta decay amplitude is observably large.

%------------------------------------------------------------------------
\section{Lorentz-invariant Description of Neutrinoless Double Beta Decay}
\label{sec:EffectiveParametrization}
%------------------------------------------------------------------------

We continue by considering the neutrinoless double beta decay rate in
a general framework, parametrizing the new physics contributions in
terms of all effective low-energy currents allowed by
Lorentz-invariance.  This parametrization has been developed in
\cite{Pas:1999fc,Pas:2000vn}.  Such an ansatz allows one to separate
the nuclear physics part of double beta decay from the underlying
particle physics model, and derive limits on arbitrary lepton number
violating theories\footnote{For another approach based on an effective operator
description compare \cite{Prezeau:2003xn}.}.  

Before discussing the general decay rate, let us recall that the 
mass mechanism of double beta decay, measures (or limits) the 
{\em effective} Majorana neutrino mass defined as:
\begin{equation}
\label{eq:defmeff}
\langle m_{\nu}\rangle = \sum_j U_{ej}^2 m_j \equiv m_{ee}.
\end{equation}
Here, the sum runs over all light neutrinos $j$ with couplings to 
the electron and a SM $W$-boson. It is straightforward to show 
that this quantity is equal to the $(ee)$ entry of the Majorana 
neutrino mass matrix in the basis where the charged lepton mass matrix 
is diagonal. For this reason $\langle m_{\nu}\rangle$ is 
often also denoted as $m_{ee}$. This contribution to neutrinoless 
double beta decay is discussed at length in the contribution by 
Rodejohann \cite{Rodejohann:2012xd} to this issue.

While the general decay rate is independent of the underlying nuclear
physics model, to extract quantitative limits values for nuclear
matrix elements are needed. Limits discussed below are derived using
matrix elements calculated in proton-neutron (pn) QRPA. For the isotope
$^{76}$Ge this matrix elements were already available in the 
literature \cite{Hirsch:1995ek,Pas:1999fc,Pas:2000vn}. For other isotopes, 
quoted in the tables below, the numerical values are taken 
from \cite{Pas:1999dr}.

Uncertainties in nuclear matrix elements are notoriously difficult to
estimate and all limits derived from double beta decay suffer from
these uncertainties. Unfortunately, despite all the efforts devoted to
the improvement of the matrix element calculations, the latest QRPA
matrix elements from the T\"ubingen group \cite{Faessler:2011rv}
differ from the shell model results revisited, for example, in
\cite{Menendez:2011zza} in many cases by factors of $\sim (2-3)$ in
case of the mass mechanism\footnote{The calculation for $^{136}$Xe
is a notable exception. Here, the latest shell model matrix element
for the mass mechanism \cite{Menendez:2011zza} 
agree with \cite{Simkovic:2007vu} within the error bars estimated 
for the QRPA calculation.}. 
Moreover, shell model matrix elements are up to now available only for
the mass mechanism. Thus uncertainties in other matrix elements,
needed in the general decay rate, are even harder to estimate.
However, for the long-range part of the amplitude, discussed in section 
(\ref{sec:longrange}) we believe that all matrix elements suffer 
from uncertainties of the same order as those found for the 
mass mechanism. 

For the short-range part of the amplitude, see section
(\ref{sec:shortrange}), no other general calculation than the one
presented in \cite{Pas:2000vn} exists. However, \cite{Faessler:2011rv}
contains matrix elements for heavy neutrino exchange and for the
short-range $R_P$-violating SUSY mechanism, which we can compare to
\cite{Pas:2000vn}. One noticeable difference is that in
\cite{Kortelainen:2007mn,Civitarese:2009zza} it was argued that the
effect of short range correlations had been overestimated in earlier
calculations. A recalculation of the nuclear matrix elements in
\cite{Simkovic:2009pp}, using the method proposed in
\cite{Kortelainen:2007mn}, indeed led to an increase of (25-40) \% in
the numerical values of the nuclear matrix elements. The latest
calculation \cite{Faessler:2011rv} also has short range matrix
elements which are larger than those in \cite{Pas:2000vn} by similar
factors. Despite these more recent calculations we will stick to the
matrix elements presented in
\cite{Hirsch:1995ek,Pas:1999fc,Pas:2000vn,Pas:1999dr}, since 
(a) no other complete calculation of matrix elements exist and 
(b) newer matrix elements in existing cases tend to be larger than 
those of the above publications, i.e. we believe that our 
limits are conservative.

Currently the most stringent bounds on neutrinoless double beta decay
come from $^{76}$Ge \cite{KlapdorKleingrothaus:2000sn} and $^{136}$Xe
\cite{Auger:2012ar}. Starting from around 2001
\cite{KlapdorKleingrothaus:2001ke,KlapdorKleingrothaus:2004wj,
KlapdorKleingrothaus:2006ff} a small part of the Heidelberg-Moscow
collaboration claimed to have observed evidence for neutrinoless
double beta decay, but this has so far not been confirmed in any other
experiment. In fact, the recent publication of the limit from
$^{136}$Xe \cite{Auger:2012ar} puts some pressure on the claim,
although, due to the uncertainty in the nuclear matrix element
calculation, $^{136}$Xe can not unequivocally rule it out yet. For
definiteness we will use the limit from $^{76}$Ge of $T_{1/2} \ge
1.9\times 10^{25}$ ys \cite{KlapdorKleingrothaus:2000sn} and the
recent result $T_{1/2} \ge1.6 \times 10^{25}$ ys for $^{136}$Xe
\cite{Auger:2012ar} for the derivation of limits. Results for these
two isotopes lead currently to very similar limits, see below\footnote{For the mass mechanism, using the nuclear matrix elements
from \cite{Muto:1989cd}, the $^{76}$Ge limit corresponds to $\emass
\lsim 0.35$ eV, while the $^{136}$Xe gives $\emass \lsim 0.34$ eV.}.

%-------------------------------------------------------------------------
\subsection{Long--Range Part}
\label{sec:longrange}
%-------------------------------------------------------------------------

%
\begin{figure}
\centering
\includegraphics[clip,width=0.7\textwidth]{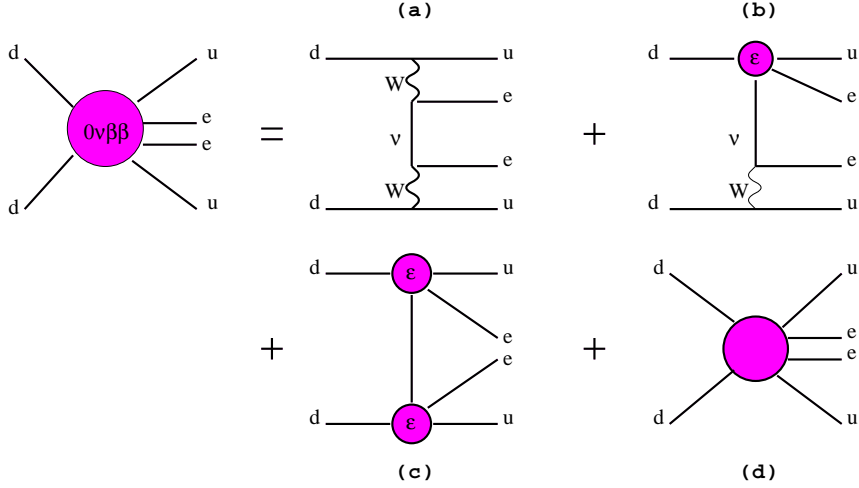}
\caption{Different contributions to the general double beta rate: 
The contributions (a) - (c) correspond to the long range part, 
the contribution (d) is the short range part (from \cite{Pas:1999fc}). 
(a) corresponds to the mass mechanism.}
\label{fig:graphs} 
\end{figure}

This subsection is essentially based on reference \cite{Pas:1999fc}. 
We consider first the long--range part of neutrinoless double beta decay
with two vertices, which are pointlike at the Fermi scale, and
exchange of a light neutrino in between. The general Lagrangian can be
written in terms of effective couplings $\epsilon^{\alpha}_{\beta}$,
which correspond to the pointlike vertices at the Fermi scale so that
Fierz rearrangement is applicable,
\begin{equation}
\label{eq:L_longrange}
{\cal L} = 
  \frac{G_F}{\sqrt{2}}\{j_{V-A}^{\mu} J^{\dagger}_{V-A,\mu} 
  + \sum_{\alpha,\beta}^{'}\epsilon_{\alpha}^{\beta}j_{\beta} J^{\dagger}_{\alpha}\},
\end{equation}
with the combinations of hadronic and leptonic Lorentz currents
$J_{\alpha}^{\dagger}=\bar{u} {\cal O}_\alpha d$ and
$j_{\beta}= \bar{e} {\cal O}_\beta \nu$ of defined helicity, respectively. The
operators ${\cal O}_{\alpha,\beta}$ are defined as
\begin{eqnarray}
\label{eq:operators}
{\cal O}_{V-A} = \gamma^{\mu}(1-\gamma_5), \qquad\qquad\qquad
{\cal O}_{V+A} = \gamma^{\mu}(1+\gamma_5),            \nonumber \\
{\cal O}_{S-P} = (1-\gamma_5),        \qquad\qquad\qquad  \qquad
{\cal O}_{S+P} = (1+\gamma_5),                        \\
{\cal O}_{T_L} = \frac{i}{2}[\gamma_{\mu},\gamma_{\nu}](1-\gamma_5), \qquad
{\cal O}_{T_R} = \frac{i}{2}[\gamma_{\mu},\gamma_{\nu}](1+\gamma_5). \nonumber
\end{eqnarray}
The prime indicates the sum runs over all contractions allowed by
Lorentz--invariance, except for $\alpha=\beta=(V-A)$. Note that all
currents have been scaled relative to the strength of the ordinary
($V-A$) interaction.

The effective Lagrangian given in eq. (\ref{eq:L_longrange})
represents the most general low-energy 4-fermion charged-current
interaction allowed by Lorentz invariance. The interpretation of the
effective couplings $\epsilon^{\alpha}_{\beta}$, however, depend on
the specific particle physics model. Nevertheless one realizes the
following general feature. Using only the SM fermion fields and
working in the Majorana basis for the neutrinos ($\nu := \nu_L +
\nu_L^C$) it is easily seen that all currents involving operators
proportional to ($1+\gamma_5$) violate lepton number by two units,
i.e. the corresponding $\epsilon^{\alpha}_{\beta}$ must also be
lepton-number violating. Such LNV $\epsilon^{\alpha}_{\beta}$ are
easily found, an example is given by $R$--parity violating supersymmetry
treated in ref. \cite{Hirsch:1995vr,Hirsch:1995ek, Pas:1998nn} and
discussed later.

The double beta decay amplitude is proportional to the time-ordered
product of two effective Lagrangians (see Fig.~\ref{fig:graphs}):
\begin{equation}
\label{eq:T_longrange}
T({\cal L}_{(1)} {\cal L}_{(2)}) = 
\frac{G_F^2}{2}
	T\{j_{V-A}J^{\dagger}_{V-A}j_{V-A}J^{\dagger}_{V-A} 
  		+ \epsilon_{\alpha}^{\beta}j_{\beta} 
  		  J^{\dagger}_{\alpha}j_{V-A} J^{\dagger}_{V-A} 
		+ \epsilon_{\alpha}^{\beta}\epsilon_{\gamma}^{\delta} 
		  j_{\beta} J^{\dagger}_{\alpha} j_{\delta} 
		  J^{\dagger}_{\gamma} \}.
\end{equation}
The first term (Fig.~\ref{fig:graphs}~(a)) corresponds
to the contribution from the Majorana neutrino mass, and the 3rd term
(Fig.~\ref{fig:graphs}~(c)), which is quadratic in
$\epsilon$ can be neglected. Only the 2nd term (Fig.~\ref{fig:graphs}~(b)) is
phenomenologically interesting. For this term one has to consider two
general cases:

\noindent
{\it 1)} The leptonic SM ($V-A$) current meets a left--handed non SM
current $j_{\beta}$ with $\beta=(S-P),T_L$. For this contribution the
neutrino propagator is
\begin{equation}
\label{eq:PropagatorLL}
P_L\frac{q^{\mu}\gamma_{\mu}+m_{\nu}}{q_{\mu}q^{\mu} - m_{\nu}^2}P_L =
\frac{m_{\nu}}{q_{\mu}q^{\mu} - m_{\nu}^2},
\end{equation}
with the usual left-- and right--handed projectors $P_{L/R}=\frac{1
  \mp \gamma_5}{2}$. This expression is proportional to the unknown
Majorana neutrino mass $m_{\nu}\lsim 0.5$ eV, for which no lower
bound exists. Therefore no limits on the corresponding parameters
$\epsilon_{\alpha}^{\beta}$ can be derived.

\noindent
{\it 2)} The leptonic SM ($V-A$) current meets a right--handed non SM
current $j_{\beta}$ with $\beta=(S+P),(V+A),T_R$. For this contribution
the neutrino propagator is
\begin{equation}
\label{eq:PropagatorLR}
P_L\frac{q^{\mu}\gamma_{\mu}+m_{\nu}}{q_{\mu}q^{\mu} - m_{\nu}^2}P_R =
\frac{q^{\mu}\gamma_{\mu}}{q_{\mu}q^{\mu} - m_{\nu}^2},
\end{equation}
which is proportional to the neutrino momentum. Since typically
$q_{\mu} \simeq p_F \simeq 100$ MeV with the nuclear Fermi
momentum $p_F$, this part of the amplitude will produce stringent
limits on corresponding $\epsilon_{\alpha}^{\beta}$.

Considering only one $\epsilon_{\alpha}^{\beta}$ at a time (evaluation
"on axis") one can now derive constraints on the effective coupling
parameters from a double beta decay half life measurement or bound,
\begin{equation}
\label{eq:T12}
[T_{1/2}^{0\nu\beta\beta}]^{-1}=|\epsilon_{\alpha}^{\beta}|^2 G_{0k} |ME|^2,
\end{equation}
where $G_{0k}$ denotes the phase space factors given in
\cite{Doi:1985dx} and $|ME|$ the nuclear matrix elements discussed
below. Note that evaluating "on axis", compared to the arbitrary
evaluation, neglects interference terms of the different
contributions. 
Numerical values of the matrix elements
${\cal M}_{T^{'}}$, ${\cal M}_{GT^{'}}$, ${\cal M}_{T^{''}}$, ${\cal
M}_{GT^{''}}$, ${\cal M}_{F^{'}}$ below, calculated in the Quasi
Particle Random Phase Approximation (pn-QRPA), are given in
Table~\ref{tab:NME}. The matrix elements for the different 
contributions are defined as follows.

\paragraph{SM meets $j_{V+A}J^{\dagger}_{V+A}$ and $j_{V+A}J^{\dagger}_{V-A}$}

These combinations of currents and the corresponding matrix elements 
have been discussed in the literature before 
\cite{Doi:1985dx, Muto:1989cd, Hirsch:1996qw}. Matrix elements 
have been calculated in different papers, and we will follow 
\cite{Hirsch:1996qw}.

%--------------------------------------------------------------------------
\paragraph{SM meets $j_{S+P}J^{\dagger}_{S+P}$ and $j_{S+P} J^{\dagger}_{S-P}$}

Using s-wave approximation for the outgoing electrons and some
assumptions according to \cite{Tomoda:1990rs, Hirsch:1995cg} one has
\begin{equation}
\label{eq:me1}
ME_{S+P}^{S+P} 
   = -ME_{S-P}^{S+P}
   = -\frac{F^{(3)}_{P}(0)}{R m_e G_A}\Big({\cal M}_{T^{'}} 
     + \frac{1}{3}{\cal M}_{GT^{'}}\Big),
\end{equation}
with the phase space factor $G_{01}$ and the Gamov-Teller and tensor
matrix elements ${\cal M}_{GT^{'}}$ and ${\cal M}_{T^{'}}$,
respectively. In addition, $R$ denotes the nuclear radius, $m_e$ 
the electron mass, $G_A \simeq 1.26$ and 
$F^{(3)}_{P}(0)= 4.41$ \cite{Adler:1975he}.

%----------------------------------------------------------------------------
\paragraph{SM meets $j_{T_{R}}J^{\dagger}_{T_{R}}$ and 
$j_{T_{R}}J^{\dagger}_{T_{L}}$}

The hadronic $T_R$ contribution is given by
\begin{equation}
\label{eq:mat02}
ME^{T_R}_{T_R} =
            - \alpha_1 \frac{2}{3} {\cal M}_{GT^{'}}
            + \alpha_1 {\cal M}_{T^{'}},
\end{equation}
with an effective nuclear form factor $\alpha_1$. For the hadronic
$T_L$ contribution in leading order of the inverse proton mass $(1/m_p)$ one finds
\begin{equation}
\label{eq:TL}
ME^{T_R}_{T_L} =
       \alpha_2 {\cal M}_{F^{'}} 
    - \alpha_3 \Big( {\cal M}_{T^{''}}+ \frac{1}{3}{\cal M}_{GT^{''}} \Big),
\end{equation}
with nuclear matrix elements ${\cal M}_{F^{'}}$, ${\cal M}_{GT^{''}}$
and ${\cal M}_{T^{''}}$ and effective nuclear form factors $\alpha_2$
and $\alpha_3$. The parameters $\alpha_i$ are defined as:
\begin{eqnarray}
\label{eq:alpiLR}
\alpha_1&=&\frac{4 T_1^{(3)}(0)G_V(1-2m_p(G_W/G_V))}{G_A^2 R m_e}, \\
\alpha_2&=&\frac{4 (2{\hat T}_2^{(3)}(0)-T_1^{(3)}(0))G_V}{G_A^2 R m_e},\\
\alpha_3&=&\frac{4 T_1^{(3)}(0)(G_P/G_A)}{G_A R^2 m_e}.
\end{eqnarray}
Here, $T_1^{(3)} =1.38$, ${\hat T}_2^{(3)} =-4.54$ \cite{Adler:1975he}, $G_P/G_A=2m_p/m_{\pi}^2$ and
$(G_W/G_V)=\frac{\mu_p-\mu_n}{2m_p}\simeq \frac{-3.7}{2m_p}$ is
obtained from the CVC hypothesis.  Numerical values for the matrix
elements and limits on the $\epsilon^{\alpha}_{\beta}$ are discussed
in Section~\ref{sec:GeneralConstraints}.

%------------------------------------------------------------------------------
\subsection{Short--Range Part}
\label{sec:shortrange}
%------------------------------------------------------------------------------

In the short range part the effective interaction can be considered as
point-like, thus the decay rate results from the following general
Lorentz invariant Lagrangian\footnote{Here we follow the essentially the calculations presented in\cite{Pas:2000vn}, see however, Table~\ref{tab:NME}.}:
\begin{eqnarray}
\label{eq:L_shortrange}
{\cal L} & =& \frac{G^2_F}{2} m_p^{-1} \{
		  \epsilon_1 J             J          j
		+ \epsilon_2 J^{\mu\nu}    J_{\mu\nu}    j
		+ \epsilon_3 J^{\mu}       J_{\mu}      j
		+ \epsilon_4 J^{\mu}       J_{\mu\nu}    j^{\nu}
		+ \epsilon_5 J^{\mu}       J           j_{\mu} \},
\end{eqnarray}
with the hadronic currents of defined
chirality $J = \overline{u}(1 \pm \gamma_5)d $, $J^{\mu} =
\overline{u} \gamma^{\mu}(1 \pm \gamma_5)d$, $J^{\mu\nu} =
\overline{u} \frac{i}{2}[\gamma^{\mu},\gamma^{\nu}](1 \pm \gamma_5)d$
and the leptonic currents $j = \overline{e}(1 \pm \gamma_5) e^C$,
$j^{\mu} = \overline{e}\gamma^{\mu}(1 \pm \gamma_5) e^C$.
In some of the cases the decay rate for the effective coupling
$\epsilon_{\alpha}$ depends also on the chirality of the currents
involved. In these cases we define $\epsilon_{\alpha} =
\epsilon^{xyz}_{\alpha}$, where $xyz = L/R, L/R, L/R$ defines the
chirality of the hadronic and leptonic currents in the order of
appearance in eq.~(\ref{eq:L_shortrange}). In the cases where it is
not necessary to distinguish the different chiralities we suppress
this additional index.

\begin{table}[t]
\centering
\begin{tabular}{cccc}
\hline
Isotope & $G_{01}$ & $G_{06}$& $G_{09}$ \\
\hline
$^{76}$Ge   & $6.40 \cdot 10^{-15}$ & $1.43 \cdot 10^{-12}$ &  $3.30 \cdot 10^{-10}$\\
$^{82}$Se   & $2.82 \cdot 10^{-14}$ & $4.77 \cdot 10^{-12}$ &  $1.32 \cdot 10^{-9}$\\
$^{100}$Mo  & $4.58 \cdot 10^{-14}$ & $7.09 \cdot 10^{-12}$ &  $1.88 \cdot 10^{-9}$\\
$^{128}$Te  & $1.83 \cdot 10^{-15}$ & $4.94 \cdot 10^{-13}$ &  $7.52 \cdot 10^{-11}$\\
$^{130}$Te  & $4.44 \cdot 10^{-14}$ & $6.89 \cdot 10^{-12}$ &  $1.55\cdot 10^{-9}$\\
$^{136}$Xe  & $4.73 \cdot 10^{-14}$ & $7.28 \cdot 10^{-12}$ &  $1.60 \cdot 10^{-9}$\\
$^{150}$Nd  & $2.10 \cdot 10^{-13}$ & $2.48 \cdot 10^{-11}$  &  $6.45 \cdot 10^{-9}$\\
\hline
\end{tabular}
\caption{Phase space factors for the general decay rate, 
numerical values taken from the calculation of \cite{Doi:1985dx}.}
\label{tab:PHSP}
\end{table}
In renormalizable theories no fundamental tensors exist. Thus the
tensor currents have to result either from Fierz rearrangements or by
integrating out heavy particles, when deriving the effective
Lagrangian from the fundamental theory and decomposing the expressions
obtained in terms of the Lorentz invariant bilinears used above,
e.g. from $\bar{u} \gamma_\mu \gamma_\nu (1+\gamma_5)d = g_{\mu\nu} J
- i J_{\mu\nu}$.  Applying the standard nuclear theory methods based
on the non-relativistic impulse approximation one derives the general
$0\nu\beta\beta$-decay half-life formula in s--wave approximation
\begin{equation}
\label{eq:t12_shortrange}
	 [T_{1/2}^{0\nu\beta\beta}]^{-1} =
			  G_{1} \left|\sum_{i=1}^3\epsilon_i {\cal M}_i\right|^2
	   	+ G_{2} \left|\sum_{i=4}^5\epsilon_i {\cal M}_i\right|^2
	   	+ G_{3} {\rm Re}\left[\left(\sum_{i=1}^3\epsilon_i {\cal M}_i\right)
			        \left(\sum_{i=4}^5\epsilon_i {\cal M}_i\right)^*\right].
\end{equation}
Here the phase space factors are
\begin{eqnarray}
\label{eqn:Phasespace_shortrange}
	G_1 = G_{01}, \ \ \
	G_2 = \frac{(m_e R)^2}{8} G_{09}, \ \ \
	G_3 = \left(\frac{m_e R}{4}\right) G_{06},
\end{eqnarray}
with $G_{0k}$ calculated in \cite{Doi:1985dx} and shown in
Table~\ref{tab:PHSP} for completeness. The nuclear matrix elements in
eq.~(\ref{eq:t12_shortrange}) are defined as\footnote{Note that the $\alpha_i$ in the long-range part 
and the $\alpha_i^{SR}$ defined here are different coefficients.}
\begin{eqnarray}
\label{eq:NME1_shortrange}
	&& {\cal M}_1 = -\alpha_1^{SR} {\cal M}_{F,N},  \ \ \ \nonumber
	   {\cal M}_2 = -\alpha_2^{SR} {\cal M}_{GT,N}, \\
	&& {\cal M}_3 = \frac{m_A^2}{m_p m_e}
	 \{{\cal M}_{GT,N} \mp \alpha_3^{SR} {\cal M}_{F,N} \}, \nonumber \\
	&& {\cal M}_4 = \pm\alpha_4^{SR} {\cal M}_{GT,N}, \ \ \
	   {\cal M}_5 = \mp\alpha_5^{SR} {\cal M}_{F,N}.
\end{eqnarray}
\begin{table}[t]
\centering
\begin{tabular}{cccccccc}
\hline
Isotope & ${\cal M}_{GT'}$ & ${\cal M}_{F'}$ & ${\cal M}_{GT''}$
& ${\cal M}_{T'}$ & ${\cal M}_{T''}$ & ${\cal M}_{GT,N}$ & ${\cal M}_{F,N}$
\\
\hline
$^{76}$Ge   & 2.95 & -0.663 & 8.78 & 0.224 & 1.33 & 0.113 & -0.0407 \\
$^{82}$Se   & 2.71 & -0.603 & 7.96 & 0.208 & 1.26 & 0.102 & -0.0360 \\
$^{100}$Mo  & 3.69 & -0.876 & 13.4 & 0.328 & 2.44  & 0.129 & -0.0489 \\
$^{116}$Cd  & 2.26 & -0.509 & 7.13 & 0.193 & 1.47 & 0.075 & -0.0271 \\
$^{128}$Te  & 3.70 & -0.814 & 13.4 & 0.331 & 2.37 & 0.119 & -0.0419 \\
$^{130}$Te  & 3.27 & -0.720 & 12.0 & 0.304 & 2.20 & 0.105 & -0.0369 \\ 
$^{136}$Xe  & 1.83 & -0.403 & 6.90 & 0.165 & 1.18 & 0.058 & -0.0203 \\ 
$^{150}$Nd  & 5.39 & -1.21  & 21.8 & 0.642 & 5.07 & 0.165 & -0.0591 \\
\hline
\end{tabular}
\caption{Nuclear matrix elements for $0\nu\beta\beta$ decay calculated
in the pn-QRPA approach. Results of different publications
\cite{Hirsch:1995ek, Pas:1999fc, Pas:2000vn,Pas:1999dr} are summarized
here.  Note that the isotope $^{150}$Nd has a sizeable deformation,
but the calculation is performed in the spherical limit. The numbers
for $^{150}$Nd might therefore be an overestimation of the true
values.}
\label{tab:NME}
\end{table}
In the last three cases contractions of hadronic currents with
different chiralities lead to different results. The negative sign in
${\cal M}_3 $ corresponds to $\epsilon_3^{LLz}$ and $\epsilon_3^{RRz}$
($J_{V\mp A}J_{V\mp A}$), the positive sign to $\epsilon_3^{LRz}$ and
$\epsilon_3^{RLz}$ ($J_{V\mp A}\-J_{V\pm A}$). The sign of ${\cal M}_4
$ is positive for the combinations $\epsilon_4^{LLL}$,
$\epsilon_4^{RRL}$, $\epsilon_4^{RLR}$, $\epsilon_4^{LRR}$ ($J_{V \mp
  A}J_{TL/TR}j_{V-A}$ and $J_{V \pm A}J_{TL/TR}j_{V+A}$). For the
combinations $\epsilon_4^{LLR}$, $\epsilon_4^{RRR}$,
$\epsilon_4^{RLL}$, $\epsilon_4^{LRL}$ ($J_{V \mp A}J_{TL/TR}j_{V+A}$
and $J_{V \pm A}J_{TL/TR}j_{V-A}$) it is negative. The sign of the
matrix element ${\cal M}_5 $ is negative for the left-handed leptonic
current ($\epsilon_5^{xyL}$) and positive for the right-handed one
($\epsilon_5^{xyR}$).
The numerical values of the standard nuclear matrix elements ${\cal M}_{F,N}$ and ${\cal M}_{GT,N}$ in eq.~(\ref{eq:NME1_shortrange}), calculated in the Quasi Particle Random Phase Approximation (pn-QRPA), are given in
Table~\ref{tab:NME}. The pre-factors $\alpha_i^{SR}$ in eq.~(\ref{eq:NME1_shortrange}) are defined as follows,
\begin{eqnarray}
\label{eq:alpha}
\alpha_1^{SR} =  
    \Big(\frac{F_S^{(3)}}{G_A}\Big)^2 \frac{m_A^2}{m_p m_e}, \ \ \
\alpha_2^{SR} = 
    8 \Big(\frac{T_1^{(3)}}{G_A}\Big)^2 \frac{m_A^2}{m_p m_e}, 
\nonumber\\
\alpha_3^{SR} =   \Big(\frac{g_V}{G_A}\Big)^2, \ \ \
\alpha_4^{SR} =   \frac{T_1^{(3)}}{G_A} \frac{m_A^2}{m_p m_e}, \ \ \
\alpha_5^{SR} =   \frac{g_V F_S^{(3)}}{G_A^2} \frac{m_A^2}{m_p m_e}.
\end{eqnarray}
The finite nucleon size is taken into account in a common way
\cite{Vergados:1980em, Vergados:1982wr} by introducing the nucleon
form factors in a dipole form
\begin{equation}
\label{eq:formeq}
\frac{g_{V,A}(q^2)}{g_{V,A}} 
= \frac{F_S(q^2)}{F_S} 
= \frac{T_1^{(3)}(q^2)}{T_1^{(3)}}=\left(1-\frac{q^2}{m_A^2}\right)^{-2},
\end{equation}
with $m_A=0.85$~GeV, $g_V= 1.0, G_A= 1.26$. The other form factor
normalizations have been calculated in ref.~\cite{Adler:1975he} within
the MIT bag model, $F_S^{(3)} = 0.48$. 

%------------------------------------------------------------------------
\subsection{General $0\nu\beta\beta$ Constraints}
\label{sec:GeneralConstraints}
%------------------------------------------------------------------------

%
\begin{table}[t]
\centering
\begin{tabular}{ccccccc}
\hline
Isotope & $|\epsilon^{V+A}_{V-A}|$ &  $|\epsilon^{V+A}_{V+A}|$ & 
          $|\epsilon^{S+P}_{S-P}|$ &  $|\epsilon^{S+P}_{S+P}|$ & 
          $|\epsilon^{TR}_{TL}|$   &  $|\epsilon^{TR}_{TR}|$   \\
\hline
$^{76}$Ge & $3.5 \cdot 10^{-9}$  & $6.2 \cdot 10^{-7}$   & 
          $1.1 \cdot 10^{-8}$    & $1.1 \cdot 10^{-8}$   & 
          $6.7 \cdot 10^{-10}$   & $1.1\cdot 10^{-9}$    \\
$^{136}$Xe & $2.8 \cdot 10^{-9}$  & $5.6 \cdot 10^{-7}$   & 
          $6.8 \cdot 10^{-9}$    & $6.8 \cdot 10^{-9}$   & 
          $4.8 \cdot 10^{-10}$   & $8.1\cdot 10^{-10}$    \\
\hline
\end{tabular}
\caption{Limits on effective long-range $B-L$ violating couplings.
These limits are derived assuming only one $\epsilon$ is different
from zero at a time. With the recent limit on the half-live for
$^{136}$Xe \cite{Auger:2012ar}, $^{136}$Xe now gives limits
competitive with or better than $^{76}$Ge.}
\label{tab:limits}
\end{table}
We list all nuclear matrix elements necessary for deriving limits for
the eight most important nuclear isotopes in Table~\ref{tab:NME}. The
long range NMEs for $^{76}$Ge had been published previously in 
\cite{Pas:1999fc}. The long-range matrix elements for other 
isotopes are from \cite{Pas:1999dr}. Note that the numerical 
values for $^{150}$Nd might overestimate the true NMEs, since the 
calculation was done in the spherical limit. With these 
matrix elements we find the limits on the different contributions 
presented in Tables~\ref{tab:limits} and \ref{tab:limits_shortrange}. 

A few comments on tables (\ref{tab:limits}) and
(\ref{tab:limits_shortrange}) might be in order. The
$j_{V+A}J^{\dagger}_{V+A}$ and $j_{V+A}J^{\dagger}_{V-A}$ part of the
amplitude has been considered within left-right symmetric models
\cite{Doi:1985dx, Muto:1989cd,Hirsch:1996qw}. In our notation, the
limits are given as $\epsilon^{V+A}_{V+A}$ and $\epsilon^{V+A}_{V-A}$.
In the notation of \cite{Doi:1985dx} these correspond to
$\langle\lambda\rangle = \epsilon^{V+A}_{V+A}$ and $\langle\eta\rangle
= \epsilon^{V+A}_{V-A}$.
Note also, that we have updated the limits with the half-live limits
from \cite{KlapdorKleingrothaus:2000sn} for $^{76}$Ge and for
$^{136}$Xe from \cite{Auger:2012ar}. All limits are derived ``on 
axis'', i.e. assuming only one non-zero contribution at a time. 

\begin{table}[t]
\centering
\begin{tabular}{ccccccc}
\hline
$^A$X & $|\epsilon_1|$ & $|\epsilon_2|$ & $|\epsilon_3^{LLz (RRz)}|$ &
$|\epsilon_3^{LRz(RLz)}|$ & $|\epsilon_4|$ & $|\epsilon_5|$ \\
\hline
$^{76}$Ge & $3.2\cdot 10^{-7}$ & $1.8 \cdot 10^{-9}$ & $2.2 \cdot 10^{-8}$ &  
 	$1.4\cdot 10^{-8}$ & $1.5 \cdot 10^{-8}$ & $1.5 \cdot 10^{-7}$ \\
$^{136}$Xe & $2.6 \cdot 10^{-7}$ & $1.4 \cdot 10^{-9}$ & $1.1 \cdot 10^{-8}$ & 
$1.7 \cdot 10^{-8}$ & $1.2 \cdot 10^{-8}$& $1.2 \cdot 10^{-7}$ \\
	\hline
\end{tabular}
\caption{Limits on effective short-range $B-L$ violating couplings.
These limits are derived assuming only one $\epsilon$ is different
from zero at a time. For $\epsilon_3$ the contractions of hadronic currents with different
  chiralities lead to different results. With the recent limit on the
  half-live for $^{136}$Xe \cite{Auger:2012ar}, $^{136}$Xe now gives
  limits competitive with or better than $^{76}$Ge.}
\label{tab:limits_shortrange}
\end{table}

%------------------------------------------------------------------------------
\section{Models of Lepton Number Violation}
\label{sec:NewPhysicsModels}
%------------------------------------------------------------------------------

In the following we discuss several prominent new physics models that incorporate lepton number violation and which lead to potentially observable rates for neutrinoless double beta decay. The list of models presented here is not intended to be exhaustive, as there is a large number of alternative schemes which produce interesting $0\nu\beta\beta$ phenomenology such as scalar bilinears \cite{Brahmachari:2002xc} or a scalar octet seesaw mechanism \cite{FileviezPerez:2009ud, Choubey:2012ux}. A large range of models is discussed in the review \cite{Rodejohann:2011mu} and in the references therein.

%------------------------------------------------------------------------------
\subsection{Left-Right Symmetry}
\label{sec:lrsymmetry}
%------------------------------------------------------------------------------

As a first example of a model incorporating a rich phenomenology of lepton number violation, we will discuss the minimal Left-Right symmetric model (LRSM) which extends the Standard Model gauge symmetry to the group SU(2)$_L~\otimes$ SU(2)$_R~\otimes$ U(1)$_{B-L}$~\cite{Pati:1974yy, Mohapatra:1974gc, Senjanovic:1975rk, Duka:1999uc}. Right-handed neutrinos are a necessary ingredient to realize this extended symmetry and are part of an SU(2)$_R$ doublet. In the LRSM, a generation of leptons is assigned to the multiplets $L_i = (\nu_i, l_i)$ with the quantum numbers $Q_{L_L} = (1/2, 0, -1)$ and $Q_{L_R} = (0, 1/2, -1)$ under SU(2)$_L~\otimes$ SU(2)$_R~\otimes$ U(1)$_{B-L}$. The Higgs sector contains a bidoublet $\phi$ and two triplets $\Delta_L$ and $\Delta_R$. The VEV $v_R$ of $\Delta_R$ breaks SU(2)$_R~\otimes$ U(1)$_{B-L}$ to U(1)$_Y$ and generates masses for the right-handed $W_R$ and $Z_R$ gauge bosons, and the heavy neutrinos. Since right-handed currents and particles have not been observed, $v_R$ has to be sufficiently large. The neutral part of the bidoublet acquires a VEV $v$ at the electroweak scale thereby breaking the SM symmetry. The LRSM can accommodate a general $6\times 6$ neutrino mass matrix in the basis $(\nu_L, \nu^c_L)^T$,
\begin{equation}
\label{eq:lrmassmatrix}
	\mathcal{M} =
	\begin{pmatrix}
		M_L & M_D \\ 
		M_D^T & M_R
	\end{pmatrix},
\end{equation}
with Majorana and Dirac mass entries of the order $M_L \approx y_M v_L$, $M_R \approx y_M v_R$ and $M_D = y_D v$. Here $y_{M,D}$ are Yukawa couplings and $v_L$ is the VEV of the left Higgs triplet, which together with the other vacuum expectation values satisfies $v_L v_R = v^2$. The mass matrix (\ref{eq:lrmassmatrix}) is diagonalized by a mixing matrix of the form
\begin{equation}
\label{eq:lrmixmatrix}
	\mathcal{U} =
	\begin{pmatrix}
		U & W \\ 
		W^T & V
	\end{pmatrix},
\end{equation}
with the $3\times 3$ block matrices $U$ and $V$ describing the mixing among the light and heavy neutrinos, respectively, whereas $W$ yields left-right mixing between the light and heavy states.

\subsubsection{Neutrinoless Double Beta Decay}
\begin{figure}[t]
\centering
\includegraphics[clip,width=0.38\textwidth]{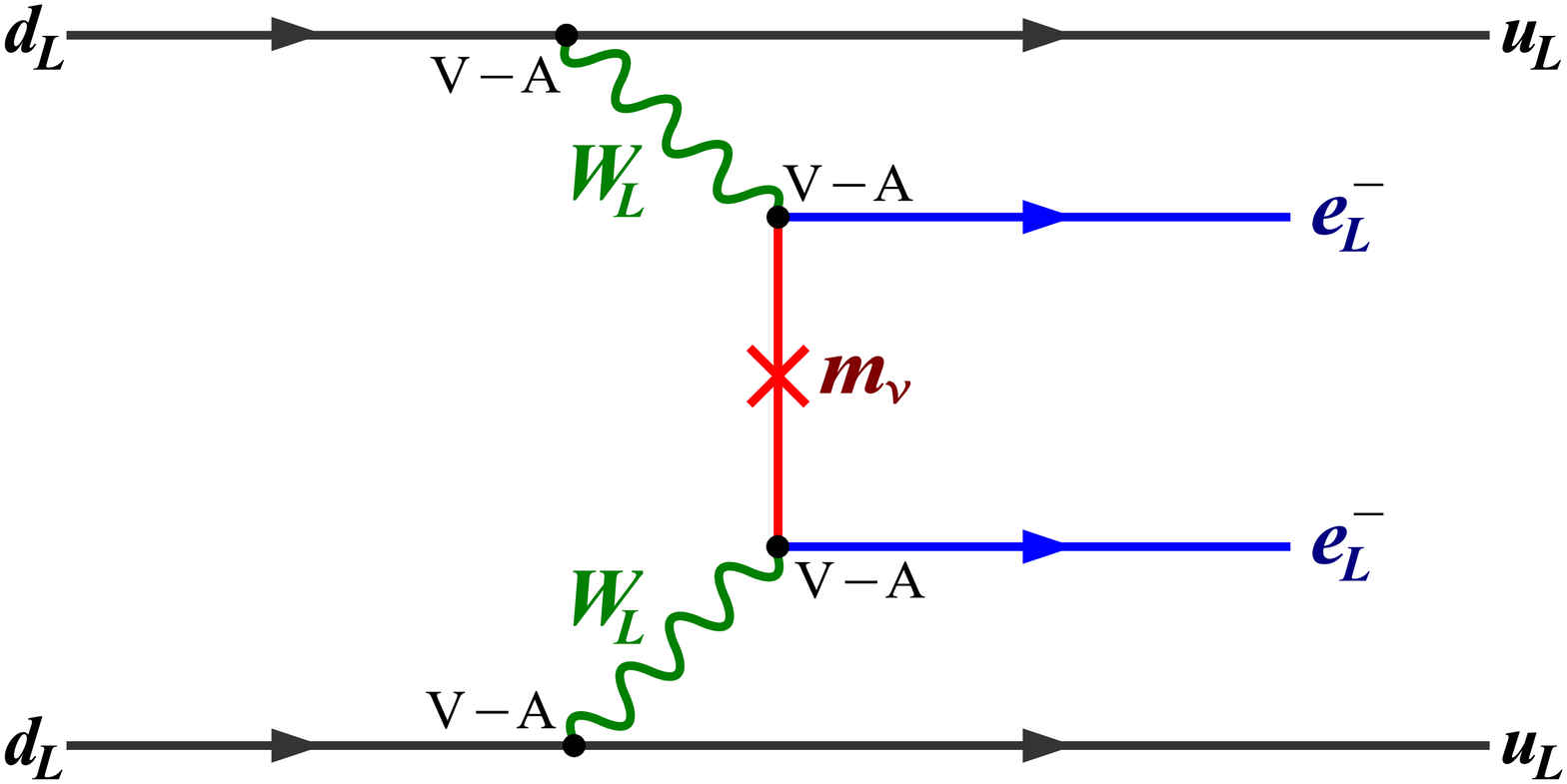}
\includegraphics[clip,width=0.38\textwidth]{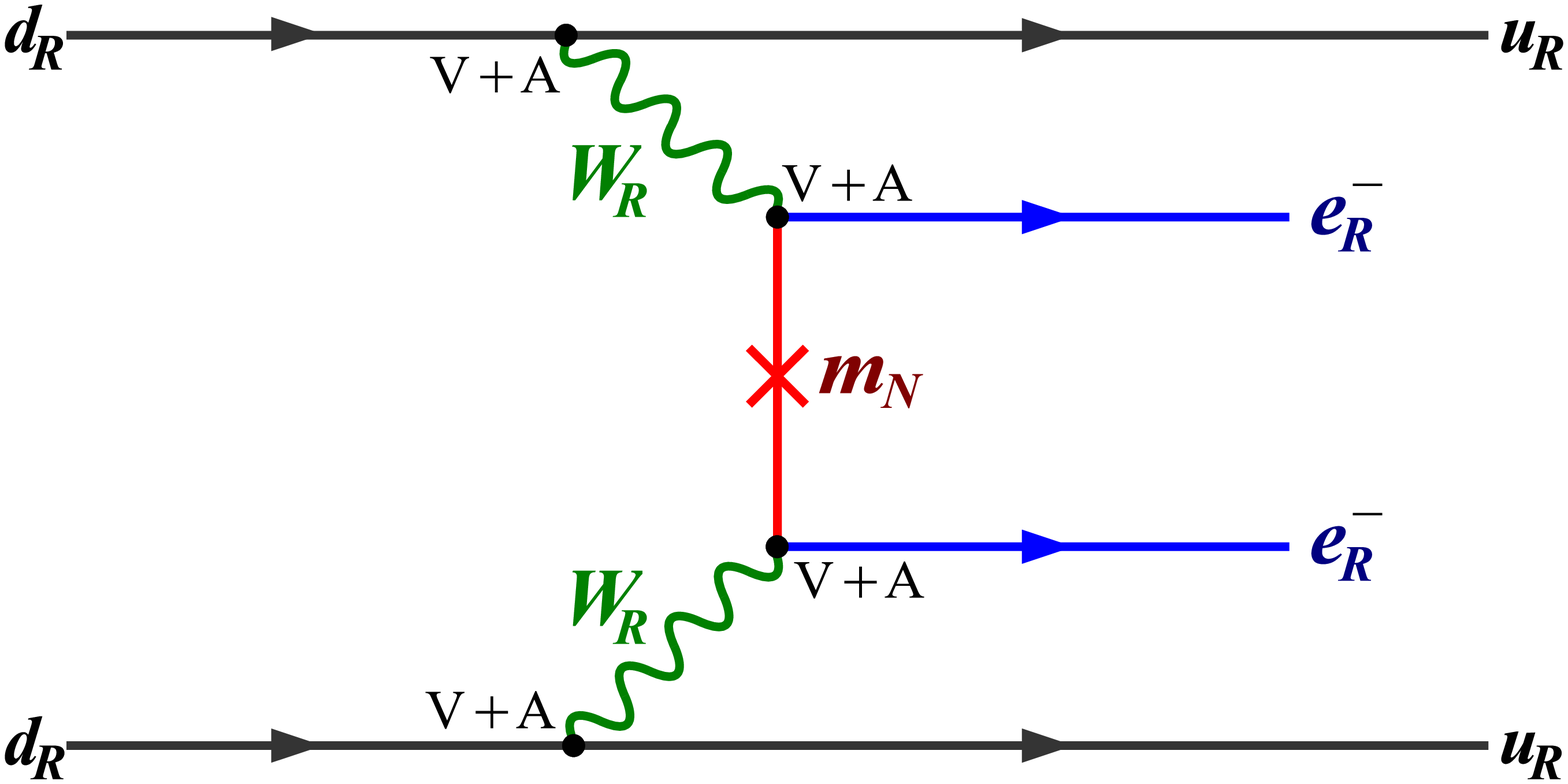}
\caption{$0\nu\beta\beta$ in the LRSM: Light (left) and heavy (right) neutrino exchange.}
\label{fig:diagramsLR_NeutrinoExchange} 
\end{figure}
\begin{figure}[t!]
\centering
\includegraphics[clip,width=0.38\textwidth]{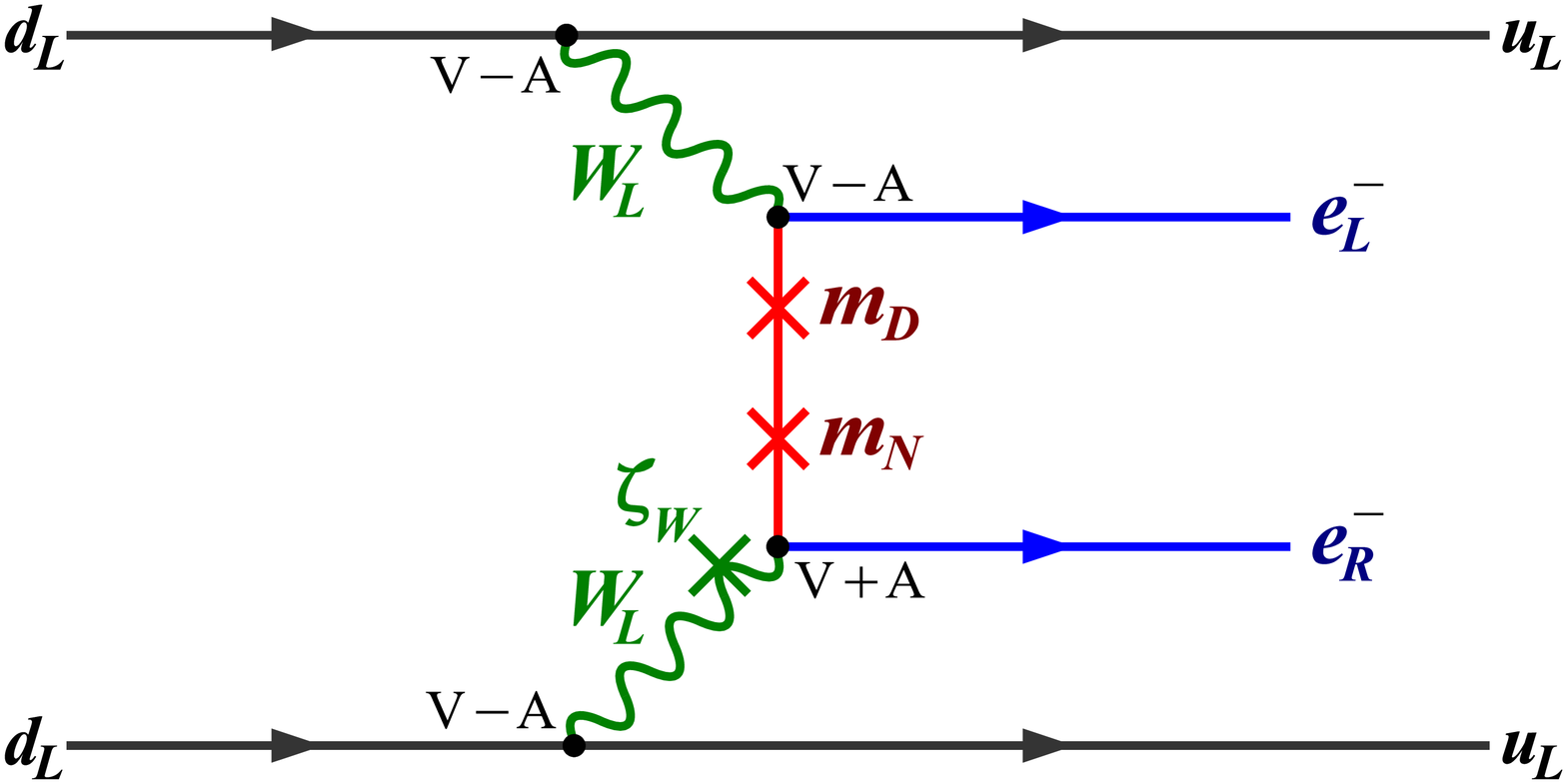}
\includegraphics[clip,width=0.38\textwidth]{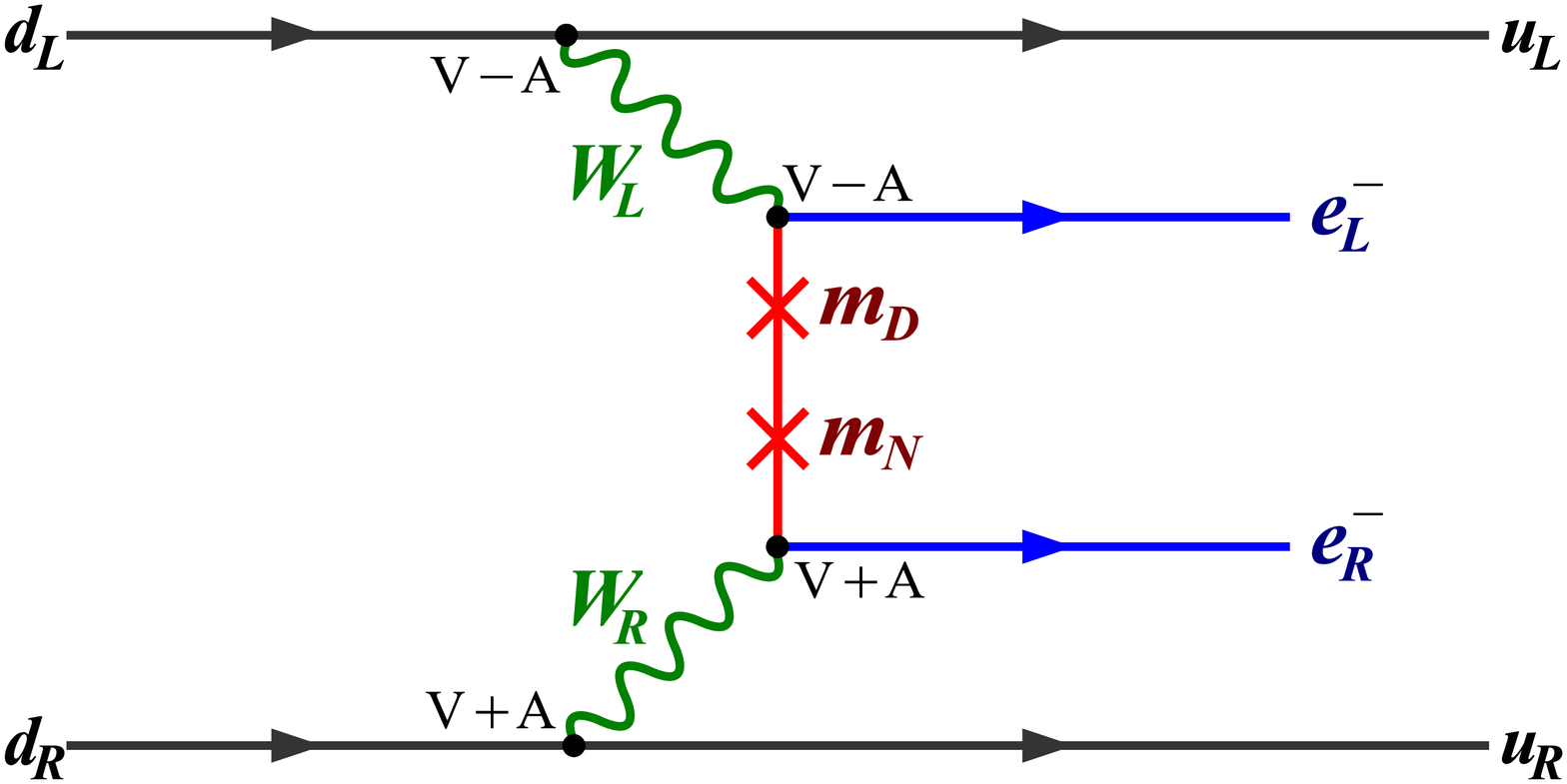}
\caption{$0\nu\beta\beta$ in the LRSM: Light neutrino exchange not proportional to the  neutrino mass, with a left-handed (left) and right-handed (right) hadronic current.}
\label{fig:diagramsLR_LR} 
\end{figure}
\begin{figure}[t!]
\centering
\includegraphics[clip,width=0.38\textwidth]{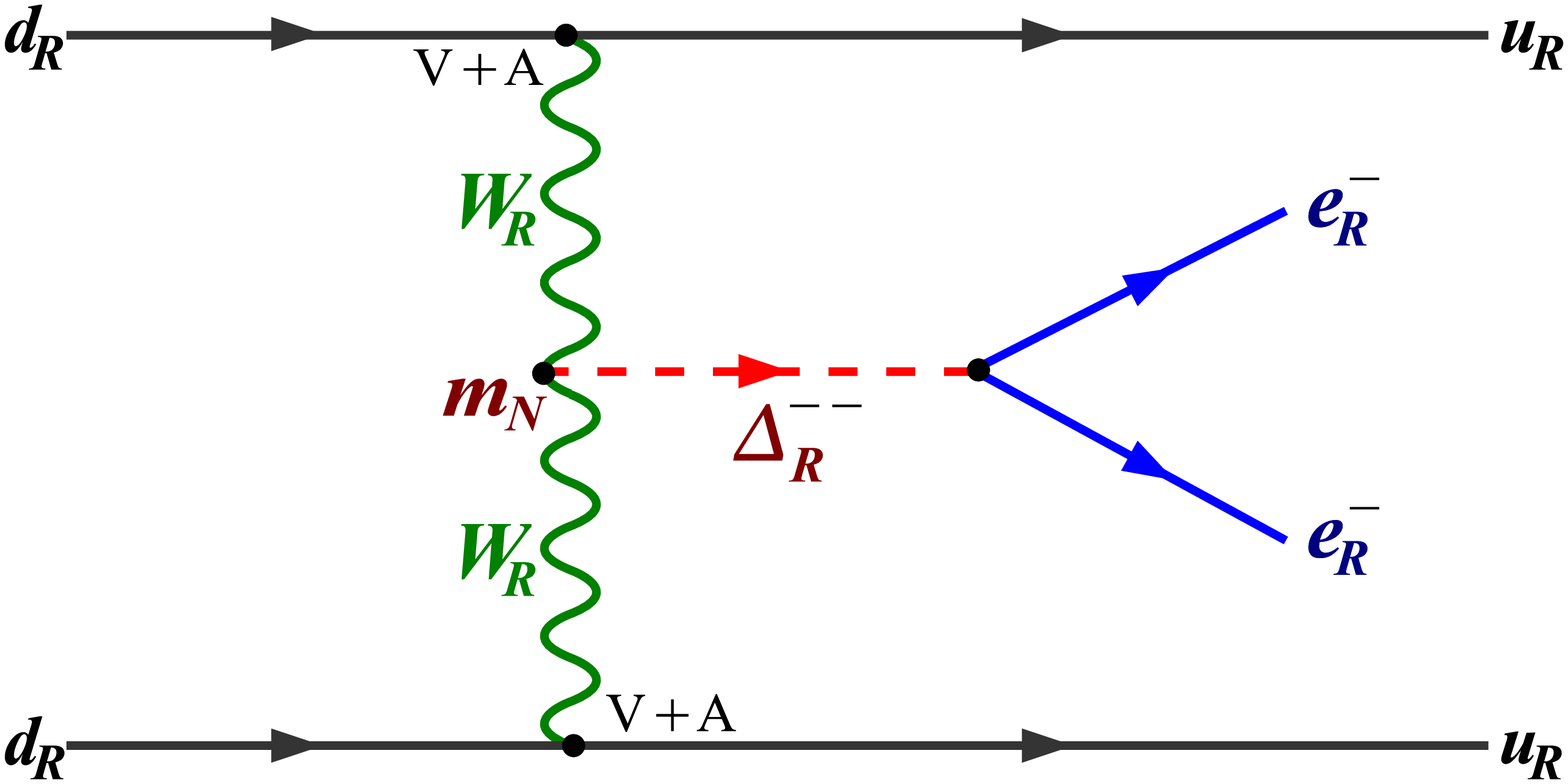}
\caption{$0\nu\beta\beta$ in the LRSM: Doubly-charged Higgs Triplet exchange.}
\label{fig:diagramsLR_Triplet} 
\end{figure}

In the LRSM, several mechanisms can contribute to $0\nu\beta\beta$ as shown in Figs.~\ref{fig:diagramsLR_NeutrinoExchange}, \ref{fig:diagramsLR_LR} and \ref{fig:diagramsLR_Triplet}. The contributions in Figs.~\ref{fig:diagramsLR_NeutrinoExchange} and \ref{fig:diagramsLR_LR} are of the same diagramatical form with the exchange of either light or heavy neutrinos as well as light and heavy $W$ bosons.
Diagram~\ref{fig:diagramsLR_NeutrinoExchange}~(left) describes the standard mechanism of light neutrino exchange, with the effective mass $m_{ee} = |\sum_i U_{ei}^2 m_{\nu_i}|$, saturating current experimental bounds if the light neutrinos are degenerate at a mass scale $m_{\nu_1} \approx m_{ee} \approx 0.3 - 0.6$~eV. Correspondingly, diagram~\ref{fig:diagramsLR_NeutrinoExchange}~(right) describes the exchange of heavy right-handed neutrinos. In the classification of Section~\ref{sec:EffectiveParametrization}, this is a realization of the short-range operator with the effective coupling $\epsilon_3^{RRz}$. 
Assuming manifest left-right symmetry, i.e. $g_R\equiv g_R$, 
in terms of the LRSM model parameters it is given by
\begin{equation}
\label{eq:epsilonN}
	\epsilon_3^{RRz} = \sum_{i=1}^3 V_{ei}^2 	
	\frac{m_p}{m_{N_i}}\frac{m_{W_L}^4}{m_{W_R}^4},
\end{equation}
and searches for $0\nu\beta\beta$ yield the limit $|\epsilon_3^{RRz}| < 1.1 \cdot 10^{-8}$ (cf. Table~\ref{tab:limits_shortrange}). 

Because of the presence of right-handed currents in the LRSM, light neutrino exchange does not necessarily require a chirality violating mass insertion. As a consequence, the outgoing electrons can have opposite chirality, and the contributions are suppressed either by the heaviness of $W_R$ or the smallness of the mixing angle $\zeta$ of the $W$ bosons as shown in Fig.~\ref{fig:diagramsLR_LR}. The coupling parameters of the corresponding effective long-range operators can be written as
\begin{equation}
\label{eq:epsilonLR}
	\epsilon_{V+A}^{V+A} = \sum_{i=1}^3 U_{ei} W_{ei} 	
	\frac{m_{W_L}^2}{m_{W_R}^2}, \qquad
	\epsilon_{V-A}^{V+A} = \sum_{i=1}^3 U_{ei} W_{ei} 	
	\tan\zeta,
\end{equation}
with the current experimental limits $|\epsilon_{V+A}^{V+A}| < 5.6 \cdot 10^{-7}$ and $|\epsilon_{V-A}^{V+A}| < 2.8 \cdot 10^{-9}$, respectively (cf. Table~\ref{tab:limits}). 
Both cases are necessarily suppressed by the left-right neutrino mixing $M_D/M_N \sim \sqrt{m_\nu/m_N}$ (the latter expression is valid for a dominant type-I seesaw mass mechanism~\cite{schechter:1981cv}) between light and heavy neutrinos.

Finally, Fig.~\ref{fig:diagramsLR_Triplet} describes the exchange of a right-handed doubly-charged triplet Higgs $\Delta_R$, which has the same effective operator structure as heavy neutrino exchange. The effective short-range coupling strength is here given as
\begin{equation}
\label{eq:epsilonDelta}
	\epsilon_3^{RRz} = \sum_{i=1}^3 V_{ei}^2 		
	\frac{m_{N_i}m_p}{m^2_{\Delta_R}}\frac{m_{W_L}^4}{m_{W_R}^4},
\end{equation}
currently limited to $|\epsilon_3^{RRz}| 
< 1.1 \times 10^{-8}$ (cf. Table~\ref{tab:limits_shortrange}). The heavy neutrino masses $m_{N_i}$ appear since in Figure~\ref{fig:diagramsLR_Triplet} the coupling of the Higgs triplet to the gauge boson is proportional to $v_R$ and the electron vertex is of Yukawa strength $(y_M)_{ee}$.

\subsubsection{Lepton Number Violation at the LHC}

\begin{figure}[t!]
\centering
\includegraphics[clip,width=0.51\textwidth]{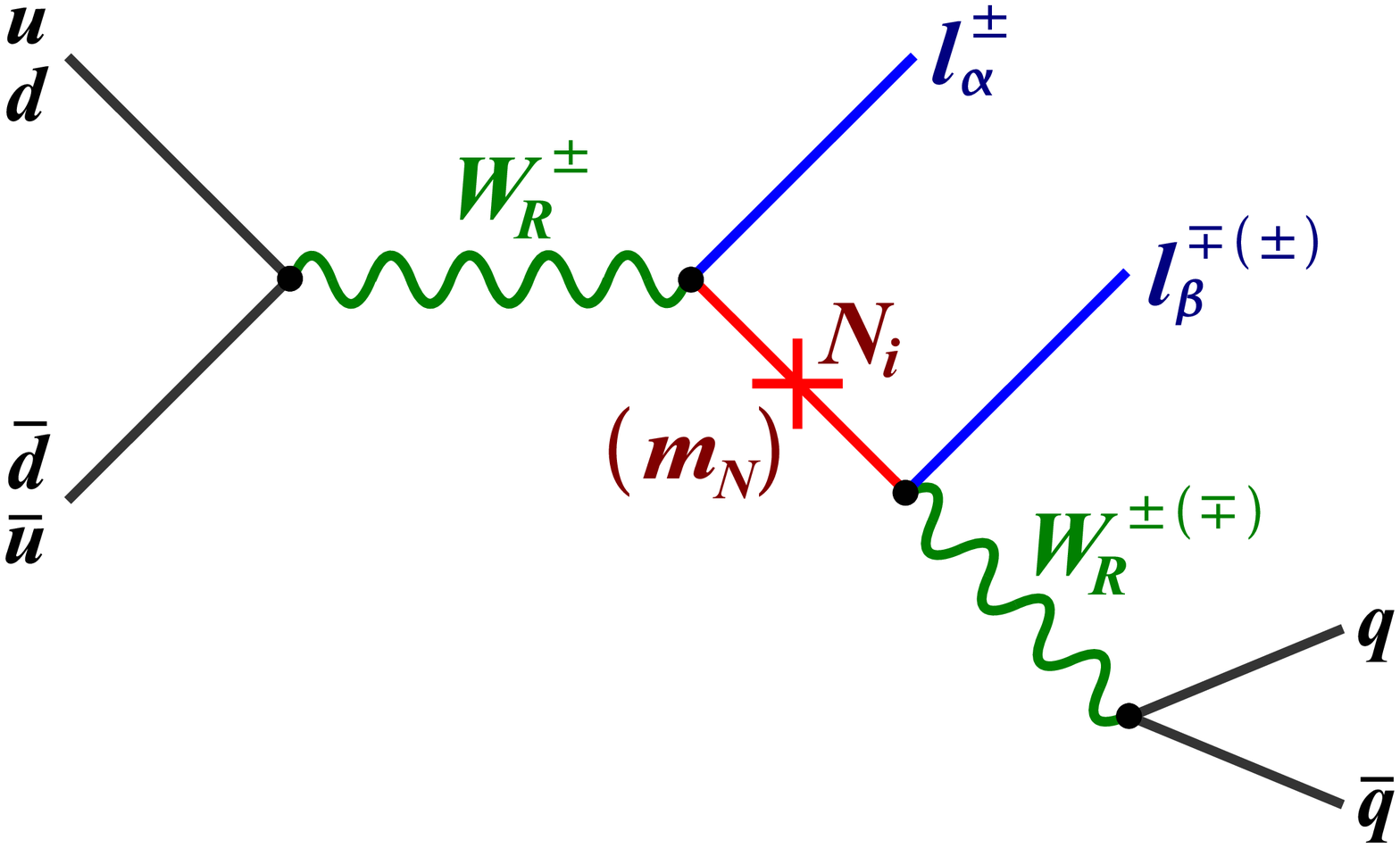}
\includegraphics[clip,width=0.42\textwidth]{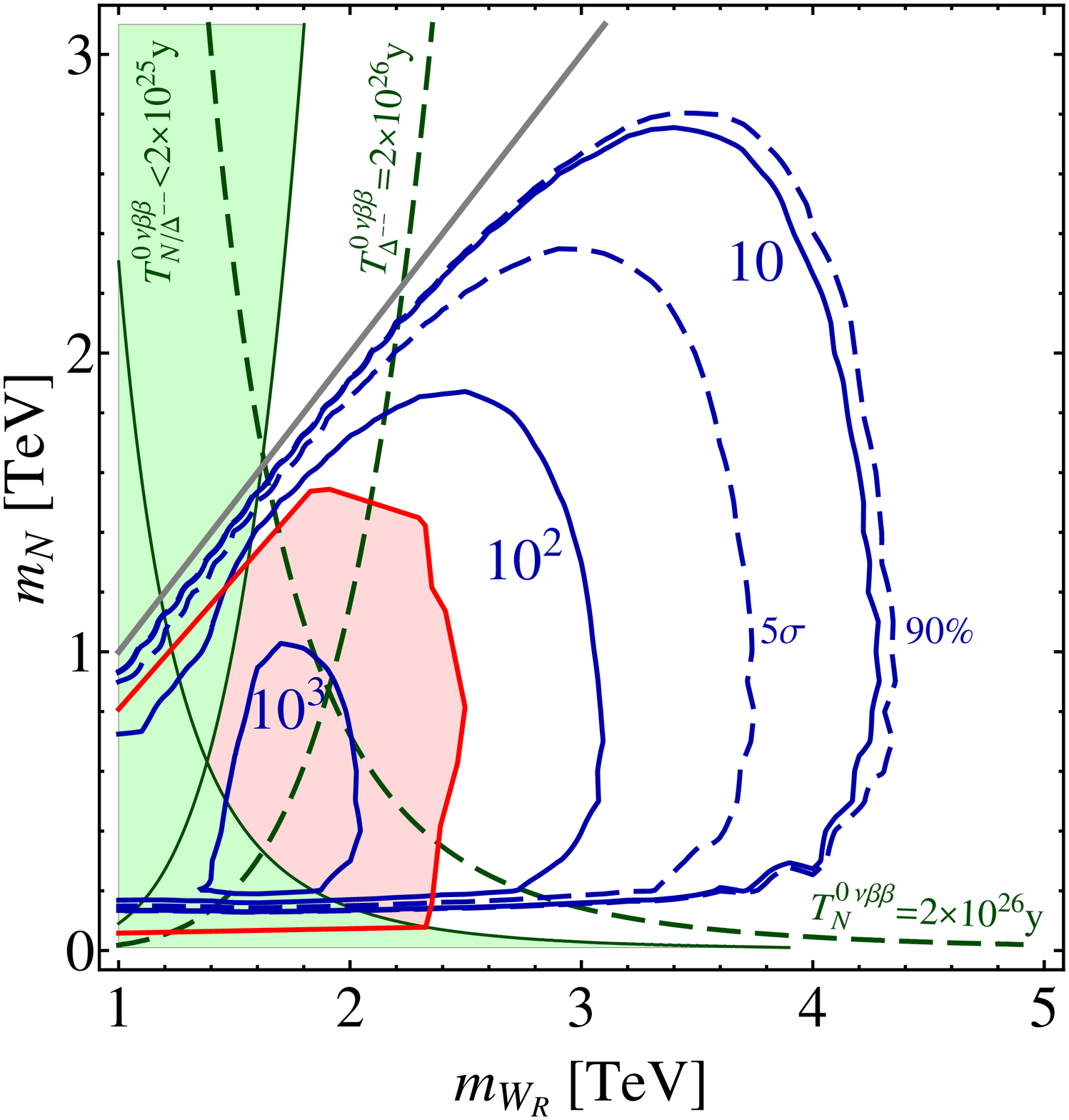}
\caption{Production and decay of a heavy right-handed neutrino with dilepton signature at hadron colliders (left). Comparison of LNV event rates at the LHC and in $0\nu\beta\beta$ experiments (right, from \cite{Das:2012ii}). The solid blue contours give the number of $e^\pm e^\pm + 2j$ events at the LHC with 14~TeV and $\mathcal{L}=30\text{ fb}^{-1}$. The dashed blue contours correspond to signal significances of $5\sigma$ and 90\%. The shaded green area denotes the parameter space excluded by $0\nu\beta\beta$ at $T^{0\nu\beta\beta}\approx 2\times 10^{25}$~years, assuming dominant doubly-charged Higgs or heavy neutrino exchange. The green dashed contours show the sensitivity of future $0\nu\beta\beta$ experiments at $T^{0\nu\beta\beta}\approx 2\times 10^{26}$~years. The red shaded area is excluded by LHC searches~\cite{CERN-PH-EP-2012-022}.}
\label{fig:diagramsLHC}
\end{figure}
In \cite{Das:2012ii}, the potential to discover number violating dilepton signals $pp\to W_R \to e^\pm \mu^{\pm,\mp} +2$~jets via a heavy right-handed neutrino at the LHC was determined, cf. Fig.~\ref{fig:diagramsLHC}~(left). Fig.~\ref{fig:diagramsLHC}~(right) compares the LHC event rates for such lepton number violation processes with the sensitivity of $0\nu\beta\beta$ experiments. The green regions and green dashed contours represent the excluded areas from $0\nu\beta\beta$ searches using nominal values for the current limit ($T^{0\nu\beta\beta} \gtrsim 2\times 10^{25}$~years) and the future sensitivity ($T^{0\nu\beta\beta} \approx 2\times 10^{26}$~years). 
In this analysis it was assumed that left-right mixing is negligible and $0\nu\beta\beta$ is either dominated by heavy neutrino or Higgs triplet
exchange. As the contribution from the standard light neutrino exchange is always present, the these results correspond to a scenario with a small effective mass $m_{ee}$. Fig.~\ref{fig:diagramsLHC}~(right) gives an example of the possible interplay between searches for lepton number violation at a high energy collider and in $0\nu\beta\beta$ experiments. 

%------------------------------------------------------------------------------
\subsection{$R$--Parity Violating Supersymmetry}
\label{sec:rpv}
%------------------------------------------------------------------------------

The MSSM (minimal supersymmetric extension of the standard model) 
assumes a discrete $Z_2$ symmetry, called $R$-=parity ($R_P$), exists. 
This symmetry guarantees the lightest supersymmetric particle 
to be stable, thus the MSSM with $R_P$ offers a dark matter candidate. 
However, from a theoretical perspective the MSSM does not offer 
any explanation as to why $R_P$ is conserved. Rather it is an 
ad hoc symmetry to avoid a phenomenological disaster. Consider 
the $R_P$ violating terms
\begin{equation}\label{R-viol} %2
W_{\rpm} = \lambda_{ijk}L_i L_j {\bar E}_k + \lambda'_{ijk}L_i Q_j {\bar D}_k
+ \epsilon_i L_i H_u + \lambda''_{ijk}{\bar U}_i {\bar D}_j {\bar D}_k,
\end{equation}
where indices $i,j, k$ label generations. The first three terms
violate lepton number, while the last one violates baryon number.  In
the presence of both types of terms the proton decays at a rate which
is many orders of magnitude above the experimental bound. However, any
discrete symmetry which eliminates either the baryon or the lepton
number violating terms is phenomenologically acceptable
\cite{dreiner:1997uz}. In fact, since the lepton number violating
terms in eq. (\ref{R-viol}) generate Majorana neutrino masses, a small
amount of $R_P$ violation could actually explain the observed neutrino
oscillation data \cite{Hirsch:2004he}. 

\subsubsection{Neutrinoless Double Beta Decay}
\label{sec:null}

In the case that $R_P$ is broken, $0\nu\beta\beta$ decay can occur through Feynman graphs
involving the exchange of superpartners as well as $\Rpv$--couplings
$\lambda^{'}$ \cite{Hirsch:1995zi, Hirsch:1995ek, Hirsch:1995cg,
Pas:1998nn}. The short-range contribution has been discussed in
\cite{Hirsch:1995zi, Hirsch:1995ek}. Here, we take the opportunity to correct some errors in the original
publication \cite{Hirsch:1995ek}, which leads to a slight change in
the numerical values of the published matrix elements: 
(i) In the coupling constants $\alpha_{A/V}^{(i)}$ of \cite{Hirsch:1995ek}, eqs.~(57)-(60), one should  replace $T_2^{(3)} \to 
{\hat T}_2^{(3)}$. (ii) In Section V of \cite{Hirsch:1995ek} five 
different nuclear matrix elements are defined. The tensor matrix 
element should be written as 
$
{\cal M}_{T'} \sim ({\bf \sigma_i \cdot {\hat r}_{ij}})
({\bf \sigma_j \cdot {\hat r}_{ij}})
- 1/3 {\bf \sigma_i \cdot \sigma_j} 
$.
And, finally (iii) there was a numerical error in the code, 
which converted Table~II to Table~III in \cite{Hirsch:1995ek}. The corrected values for the nuclear matrix element ${\cal M}_{\tilde q}$ are given in Table~\ref{tab:FinTab}. As in the original
paper, $-{\cal M}_{\tilde q}$ is given, to account for a relative sign
in the neutrino mass mechanism with respect to the definitions used in
the formalism of Doi, Kotani and Takasugi \cite{Doi:1985dx}. Note 
that the $R_P$ violating SUSY mechanism 
depends on a combination of different short-range 
matrix elements discussed in section (\ref{sec:shortrange}), since 
in the amplitude both $JJj$ and $J^{\mu\nu}J_{\mu\nu}j$ currents 
appear.

\begin{table}[t]
\centering
\vspace*{5mm}
\begin{tabular}{ccccccccc}
\hline
$^{A}$Y & $^{76}$Ge & $^{82}$Se & $^{100}$Mo & $^{116}$Cd & $^{128}$Te & 
$^{130}$Te & $^{136}$Xe & $^{150}$Nd \\
\hline
A) ${\cal M}_{\tilde q}$ & 222 & 199 & 265 & 150 & 242 & 214 & 116 & 353 \\
B) ${\cal M}_{\tilde q}$ & 281 & 251 & 331 & 190 & 301 & 266 & 145 & 432 \\
\hline
\end{tabular}
\caption{Nuclear matrix elements for short-range SUSY $0\nu\beta\beta$ decay. 
Shown are ${\cal M}_{\tilde q}$ for the two sets of input values of 
coefficients for the $\alpha^{(i)}$ of Table~1 in \cite{Hirsch:1995ek}, 
corrected for the errors discussed in the text.}
\label{tab:FinTab}
\end{table}

Ref. \cite{Hirsch:1995ek} used the half-live limit from the
Heidelberg-Moscow collaboration available at that time
\cite{Balysh:1995zc}, which was later superseded by the more
stringent value of \cite{KlapdorKleingrothaus:2000sn}.  The change in
the matrix element and the update in the half-life combined leads to a
slightly more stringent limit on $\lambda_{111}^{'}$ given by
\begin{equation}
	\lambda_{111}^{'}\leq 
		2.6\cdot 10^{-4}\Big(\frac{m_{\tilde{q}}}{100 \text{ GeV}} \Big)^2
		\Big(\frac{m_{\tilde{g}}}{100 \text{ GeV}} \Big)^{1/2},
\end{equation}
for $m_{\tilde{d}_{R}}=m_{\tilde{u}_{L}}$. For comparison, from
$^{136}$Xe data \cite{Auger:2012ar} one gets currently
$\lambda_{111}^{'}\leq 2.7\cdot 10^{-4}$. Note, that recently
\cite{Faessler:2011rv} published sets of matrix elements for $R_P$
violating double beta decay for the pion-exchange mechanism, which
are, depending on the choice of nucleon-nucleon interaction, model
space and value of $G_A$, between a factor of $x\sim (2-3)$ larger than the
matrix elements of Table~\ref{tab:FinTab}. These would lead to limits on $\lambda_{111}^{'}$ which are stronger by a 
corresponding factor of $\sqrt{x}$.

$0\nu\beta\beta$ decay is not only sensitive to
$\lambda_{111}^{'}$. Taking into account the fact that the SUSY
partners of the left- and right--handed quark states can mix with each
other, new diagrams appear in which the neutrino-mediated double beta
decay is accompanied by SUSY exchange in the vertices
\cite{Babu:1995vh, Hirsch:1995cg, Pas:1998nn}. A calculation of 
previously neglected tensor contributions to the decay rate allows to
derive improved limits on different combinations of $\lambda^{'}$
\cite{Pas:1998nn}. Assuming the supersymmetric mass parameters of
order 100 GeV, the half life limit of the Heidelberg--Moscow
Experiment implies: $\lambda_{113}^{'} \lambda_{131}^{'}\leq 3 \cdot
10^{-8}$, $\lambda_{112}^{'} \lambda_{121}^{'}\leq 1 \cdot 10^{-6}$.

\subsubsection{Lepton Number Violation at the LHC}

Similar to the situation in left-right symmetric models, 
also in $R$--parity violating SUSY scenarios interesting
and complementary information can be obtained from the 
neutrinoless double beta decay
analogue at the LHC, namely resonant single selectron production 
with two like sign electrons
in the final state \cite{Allanach:2009iv,Allanach:2009xx}.
The color and spin-averaged parton total
cross section of a single slepton production is given 
by~\cite{Dimopoulos:1988jw}
\begin{equation}
\hat{\sigma} = \frac{\pi}{12\hat{s}}|\lambda'_{111}|^2
\delta\left(1-\frac{m^2_{\tilde{l}}}{\hat{s}}\right),
\end{equation}
where $\hat{s}$ is the partonic center of mass energy,
$m_{\tilde{l}}$ is the mass of the resonant slepton, and finite width effects
have been neglected.
Considering effects from
parton distribution functions, 
to a good approximation
the total cross section scales like
$\sigma (pp \rightarrow \tilde l) \propto |\lambda'_{111}|^2 / m_{\tilde l}^3 $
with the slepton mass
in the parameter region of interest.

Thus
for small slepton masses, the stringent bound 
from $0\nu\beta\beta$ decay makes this process unobservable at the LHC.
However, the
bound on $\lambda'_{111}$ originating 
from the non-observation of $0\nu\beta\beta$ decay scales 
with the slepton mass like 
$\sigma< c \Lambda_{SUSY}^2$ where $c$ is a constant, so that for
higher values of the SUSY masses, larger cross-sections may be
allowed as much a larger $\lambda'_{111}$ is no longer excluded. 
It is this possibility
that can be exploited at the LHC. 

\begin{figure}[t!]
\centering
\includegraphics[clip,width=0.38\textwidth]{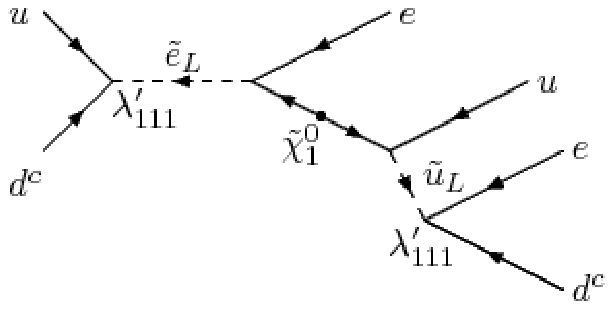}
\includegraphics[clip,width=0.5\textwidth]{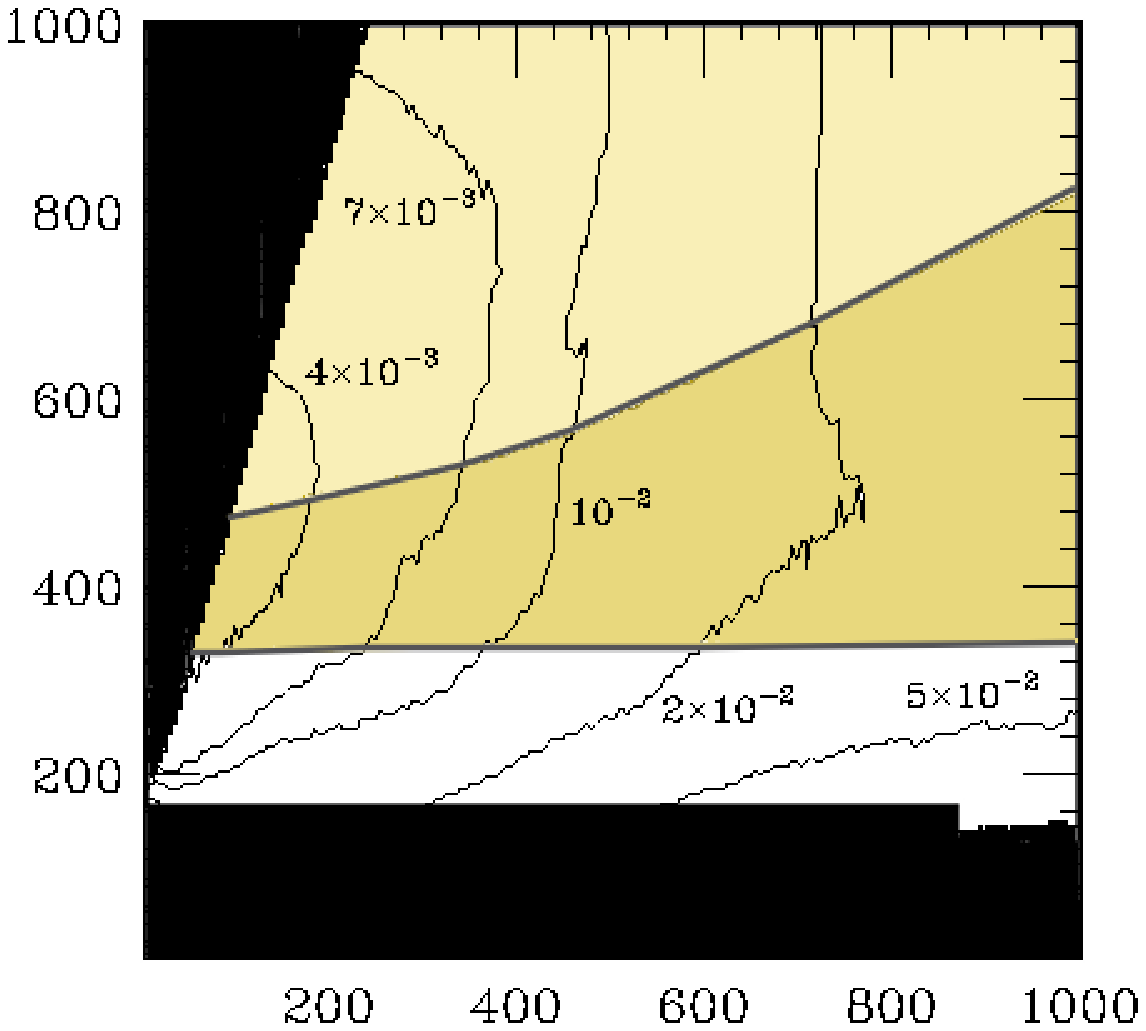}
\caption{Production and decay of a single selectron in $R$--parity 
violating models (left, from 
\cite{Allanach:2009iv}) .
Region in the 
mSUGRA parameter space (vertical axis: $M_{1/2}$ in GeV, horizontal axis:
$M_0$ in GeV)  in which
  single slepton production may be observed at the LHC for $\tan \beta=10$, 
  $A_0=0$ and 10fb$^{-1}$ of integrated luminosity. 
  In the top left-hand black triangle, the stau is the LSP, a case not 
  discussed in \cite{Allanach:2009iv}. 
The black region at the bottom is ruled out by direct
  search constraints. The labeled contours are taken from
  Ref.~\cite{Dreiner:2000vf}, and indicate the search reach given by the 
corresponding
  value of $\lambda'_{111}$. The white, dark-shaded and light-shaded regions
  demonstrate that observation of single slepton production at the 
$5\sigma$ level would imply     
$T^{0\nu\beta\beta}_{1/2} < 1.9\cdot 10^{25} \rm{yrs}$,  
$100> T^{0\nu\beta\beta}_{1/2}/10^{25} \rm{yrs} > 1.9$ and
    $T^{0\nu\beta\beta}_{1/2} > 1 \times 10^{27} \rm{yrs}$, respectively
    (right, from \cite{Allanach:2009iv}).
 }
\label{fig:diagramsRPV-LHC} 
\end{figure}

In \cite{Allanach:2009iv,Allanach:2009xx} it has been shown that much of the
parameter space allowed by $0\nu\beta\beta$ decay in simple models of 
supersymmetry breaking actually predicts observable single slepton 
production at the
LHC\@ (compare Fig.~\ref{fig:diagramsRPV-LHC}). Moreover, if
the next generation of experiments observe $0\nu\beta\beta$ decay, 
the LHC has a
very good chance of observing single slepton production with only 10
fb$^{-1}$ of integrated luminosity, assuming that $0\nu\beta\beta$ decay 
is induced by a
$\lambda'_{111}$ coupling. On the other hand, non-observation 
of single slepton production could then discriminate against 
the $\lambda'_{111}$ mechanism.
In general, both Majorana neutrino masses and
$\lambda'_{111}$ could contribute 
simultaneously and non-negligibly to $0\nu\beta\beta$ decay. 
In this case detailed LHC measurements of the kinematics 
in single slepton production
could constrain  
the SUSY parameters, and the total cross-section could then give information
about the size of $|\lambda'_{111}|$. 
In principle and depending on nuclear matrix element uncertainties
the LHC information could be combined to
predict an associated 
inverse double beta decay half life coming from $\lambda'_{111}$, which could
be compared with the experimental half life measurement  in order to see
if additional contributions were necessary.

%------------------------------------------------------------------------------
\subsection{Leptoquarks}
\label{sec:leptoquarks}
%------------------------------------------------------------------------------

\begin{figure}
\centering
\includegraphics[clip,width=0.3\textwidth]{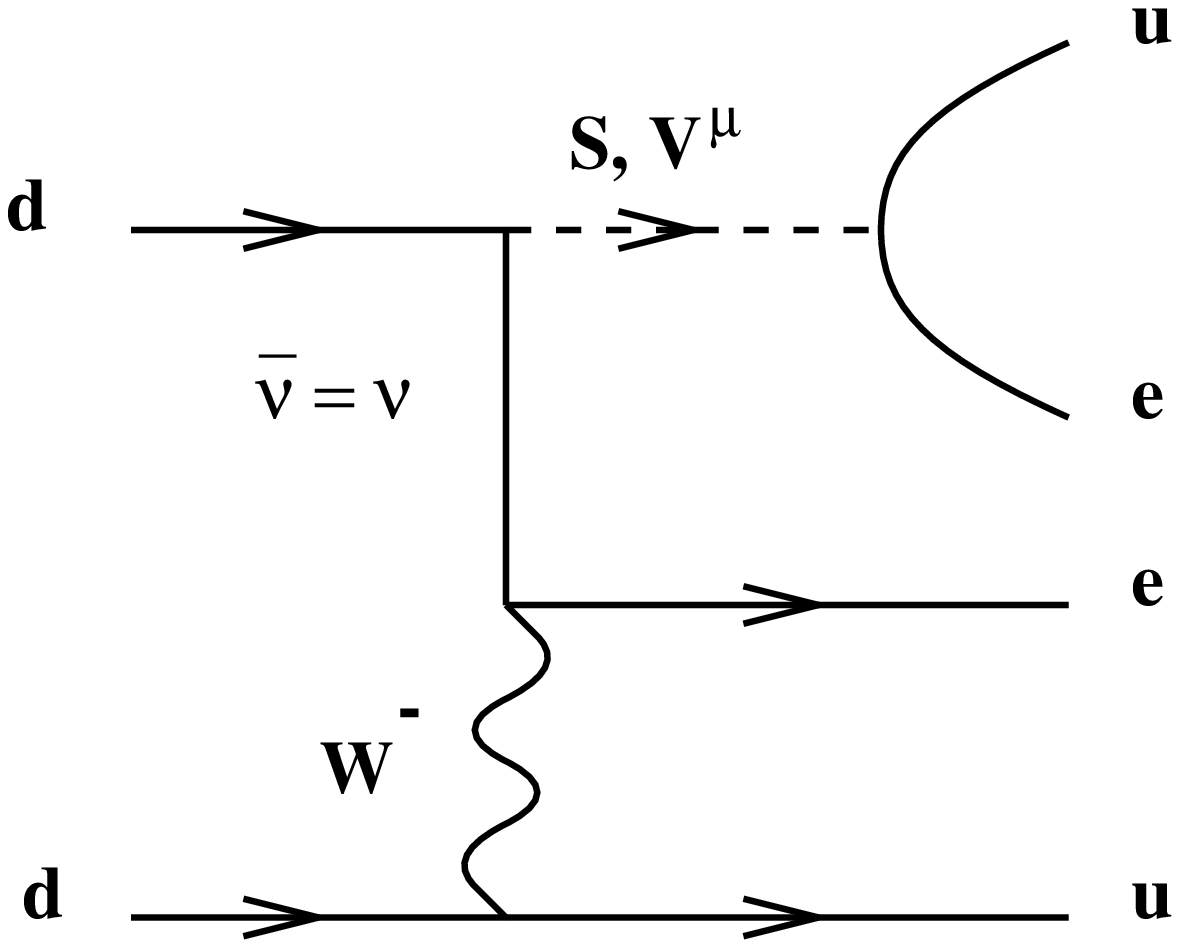}
\includegraphics[clip,width=0.3\textwidth]{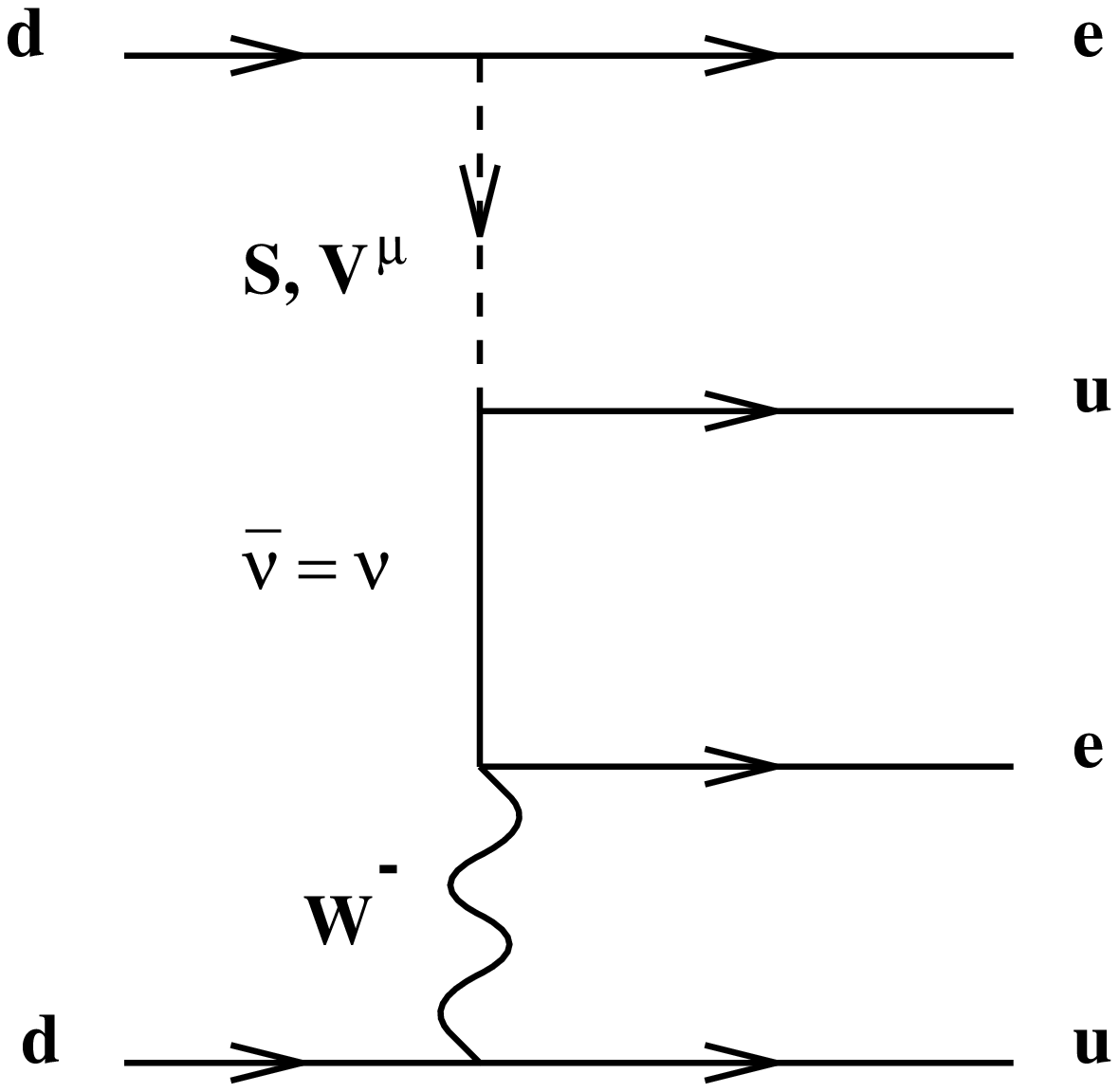}
\caption{\label{fig:bblq}Leptoquark diagrams for neutrinoless 
double beta decay.}
\end{figure}

Leptoquarks (LQs) are hypothetical scalar or vector particles coupling 
to both leptons and quarks. They appear most prominently in grand 
unified theories, but also in extended Technicolor or Compositeness 
models. LQs which conserve baryon number can be relatively light 
\cite{Buchmuller:1986zs}, possibly within reach of accelerator experiments. 
Also low-energy precision measurements can give limits on LQ properties, 
for a detailed list on constraints from non-accelerator 
searches see, for example \cite{Davidson:1993qk} and \cite{Nakamura:2010zzi}. 
The mixing of different LQ multiplets by a possible leptoquark--Higgs
coupling \cite{Hirsch:1996qy} can lead to a contribution to
$0\nu\beta\beta$ decay, if these couplings violate lepton number 
\cite{Hirsch:1996ye}. Diagrams involving LQs and standard model 
weak current interactions can be generated, see Fig.~\ref{fig:bblq}. 
These diagrams are of the long range type and due to the chirality 
violating LQ interaction gain a $\pslash$-enhancement in the 
double beta decay amplitude.
Combined with a lower limit on the $\znbb$ decay half--life bounds on
effective couplings can be derived \cite{Hirsch:1996ye}. Assuming only
one lepton number violating $\Delta L=2$ LQ--Higgs coupling unequal to
zero and the leptoquark masses not too different, one can derive from
this limit a bound on the LQ--Higgs coupling,
\begin{equation}
	Y_{LQ-\text{Higgs}} = \text{few}\cdot 10^{-6},
\end{equation}
for LQ masses of the order of ${\cal O}(200 \text{ GeV})$. Such lepton number violating LQ-Higgs couplings also lead to 
non-zero neutrino masses at the 1-loop level \cite{AristizabalSierra:2007nf}.

%------------------------------------------------------------------------------
\subsection{Extra Dimensions}
\label{sec:extradimensions}
%------------------------------------------------------------------------------

Theories with
large   compact extra dimensions   of~TeV size~\cite{ArkaniHamed:1998rs, Antoniadis:1998ig, Dienes:1998vh, Dienes:1998vg, Randall:1999ee, Antoniadis:1990ew, Lykken:1996fj, Horava:1995qa, Horava:1996ma} have  enriched
dramatically  the perspectives of the search for physics beyond  the
Standard  Model.  Among   the  possible   higher-dimensional
realizations,  sterile    neutrinos  propagating   in such  extra
dimensions~\cite{Dienes:1998sb, ArkaniHamed:1998vp, Pilaftsis:1999jk, Ioannisian:1999cw}  may  provide interesting  alternatives for generating  
the observed  light neutrino
masses.  Conversely, detailed  experimental studies of neutrino
properties may even shed light on the geometry and shape of the new
dimensions.

The minimal  higher-dimensional
framework of lepton number violation considers a 5-dimensional theory 
compactified  on a $S^1/Z_2$ orbifold,
in  which only one 5-dimensional
(bulk) sterile neutrino is added  to the field content of  the SM
\cite{Bhattacharyya:2002vf}.   
In this minimal model, the SM   fields are localized on a   3+1-dimensional
Minkowski subspace, also termed 3-brane. This model naturally generates 
small neutrino masses via a higher-dimensional version of the seesaw 
mechanism \cite{Dienes:1998sb}.
With respect to neutrinoless double beta decay an interesting feature of such 
extra-dimensional models is that the excitations of the sterile neutrino 
in the compact extra dimensions, a so-called Kaluza-Klein tower of states, 
contributes to the decay rate. As the masses of the exchanged Kaluza-Klein
excitations range from small masses giving rise to long range contributions
over the 100~MeV region up to large masses with short range contributions,
such scenarios constitute a special case which is not described by the
effective operator parametrization in Section~3. 

Other
studies on neutrinoless  double beta decay were performed  within the
context  of   higher-dimensional models   that assume a shining
mechanism from a distant brane~\cite{Mohapatra:2000px} and of 
theories with wrapped
geometric       space~\cite{Huber:2002gp}. In the first model  
$0\nu\beta\beta$ decay  is accompanied with the emission of Majorons
\cite{Mohapatra:2000px}, while 
the model presented in~\cite{Huber:2002gp} 
predicts double beta decay at a rate 
being too small to be observable signal in running experiments.
Another approach motivated by \cite{ArkaniHamed:1998vp}
which does not consider the effect of Kaluza-Klein states is described in
\cite{Gozdz:2006iz}.

\begin{figure}[t]
\centering
\includegraphics[clip,width=0.4\textwidth]{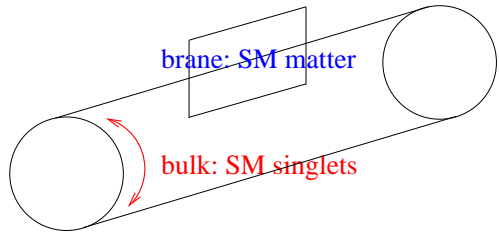}
\caption{The SM matter is localized on a 3-brane, while
the sterile singlet neutrinos are allowed to propagate in the bulk.
This framework naturally generates small neutrino masses.}
\label{extradim} 
\end{figure}

The 
minimal higher-dimensional scenario in 
\cite{Bhattacharyya:2002vf}
assumes    that singlet neutrinos being
neutral under   the SU(2)$_L\otimes$U(1)$_Y$ gauge  group  can  freely
propagate   in  a    higher-dimensional   space  of   $[1+(3+\delta)]$
dimensions, the so-called bulk, whereas all SM particles are localized
on the $(3+1)$-dimensional brane.
If the brane were located at the one of the two orbifold fixed points,
the lepton number violating operators would be absent as
a consequence  of the $Z_2$ discrete symmetry.   However, if the brane
is shifted  by an amount   $a\ne 0$, these operators are no longer
absent and this breaking
of  lepton  number can lead  to   observable effects  in  neutrinoless
double beta decay experiments.

One   major problem of such   higher-dimensional theories is  their
generic   prediction  of a   KK  neutrino  spectrum  of  approximately
degenerate states with opposite  CP parities that lead to  extremely
suppressed values for the  effective Majorana-neutrino mass $\langle m_\nu
\rangle$. However, the  KK neutrinos can  couple  to the $W$
bosons with unequal strength, thus avoiding the CP-parity
cancellations in the  $0\nu\beta\beta$-decay amplitude. 
The brane-shifting parameter  $a$   can then  be  determined from  the
requirement that the effective Majorana mass $\langle m_\nu \rangle$ is in
the observable range. To achieve this,  $1/a$ has to 
be constrained to be  
larger  than the  typical Fermi nuclear
momentum $q_F  = 100~{\rm  MeV}$  and much  smaller  than the  quantum
gravity     scale    $M_F$,   or     equivalently    $1/M_F    \ll   a
\stackrel{<}{{}_\sim} 1/q_F$.

The masses of the Kaluza-Klein states are obtained then by diagonalizing
the infinitely dimensional Kaluza-Klein mass matrix and result
approximately as
\begin{equation}
  \label{mn}
m_{(n)}\ \approx\  \frac{n}{R}\ + \: \varepsilon\; ,
\end{equation}
where $n$ is the index denoting the Kaluza-Klein excitation, 
$R$ is the radius of the extra
dimension
and $\varepsilon$ is the smallest diagonal entry in the neutrino mass matrix.
The  mixing-matrix elements
$B_{e\nu}$ and $B_{e,n}$ follow then as \cite{Dienes:1998sb}:
\begin{equation}
B_{e\nu}  =    \frac{1}{1 + \pi^2 m^2 R^2 +
\frac{m^2_\nu}{m^2}}, \quad
B_{e,n} \simeq\frac{m^2 \cos^2(\, \frac{na}{R} - \phi_h\,)}
{(\,\frac{n}{R} + \varepsilon\,)^2},
\label{ben}
\end{equation}
where $a$ is the brane shift parameter and $\phi_h$ is a function of $a$, $R$ and the Yukawa couplings. 
The $0\nu\beta\beta$-decay amplitude ${\cal
T}_{0\nu\beta\beta}$ results as
\begin{equation}
  \label{M2beta}
{\cal T}_{0\nu\beta\beta}\ =\ \frac{\langle m_\nu \rangle}{m_e}\ 
                         {\cal M}_{\rm GTF} (m_\nu )\; ,
\end{equation}
where the combination of nuclear matrix elements 
${\cal M}_{\rm  GTF} = {\cal M}_{\rm GT}  - {\cal M}_{\rm F}$ 
sensitively  depends   on the  mass  of the
exchanged KK neutrino. As soon as the
exchanged KK-neutrino mass $m_{(n)}$ is comparable  or larger than the
characteristic  Fermi  nuclear   momentum $q_F  \approx 100$~MeV,  the
nuclear   matrix element    ${\cal     M}_{\rm  GTF}$  decreases    as
$1/m^2_{(n)}$.    The  general     expression  for    the    effective
Majorana-neutrino mass $\langle  m_\nu \rangle$ in eq.~(\ref{M2beta}) is given
by
\begin{equation}
  \label{effMajmass}
\langle m_\nu \rangle\ =\ \frac{1}{{\cal M}_{\rm GTF} (m_\nu) }\
\sum_{n = -\infty}^{\infty}\, B^2_{e,n}\,m_{(n)}\, 
\Big[\, {\cal M}_{\rm GTF} (m_{(n)})\: 
-\: {\cal M}_{\rm GTF} (m_\nu )\, \Big]\; .\quad
\end{equation}
Here the  first term describes the genuine higher-dimensional
effect of KK-neutrino exchanges, while the second term is the standard
contribution  of  the  light   neutrino  $\nu$. 
The dependence of the nuclear matrix element
${\cal  M}_{\rm GTF}$  on the  KK-neutrino masses  $m_{(n)}$  here has been
included in the double beta observable  $\langle m_\nu  
\rangle$.   This leads to predictions for  $\langle m_\nu \rangle$ that depend on
the  double beta  emitter isotope  used in the experiment.   However, the
difference in the predictions  is typically too small 
to work  as a smoking  gun for the
extra-dimensional mechanism of  $0\nu\beta\beta$ decay.
A particularly  interesting   property of this model is that  the  effective
Majorana-neutrino mass $\langle  m_\nu \rangle$ 
is not bounded from above by the mass eigenvalues of the light neutrinos:
It can  be  
close to the experimental limit even for 
an infinitesimal lightest neutrino mass $m_{\nu_1}$ which
constitutes
a  rather  unique feature of such higher-dimensional brane-shifted
scenarios.

%------------------------------------------------------------------------------
\subsection{Majorons}
\label{sec:majorons}
%------------------------------------------------------------------------------

In many theories beyond the Standard model in addition to the 
neutrinoless double beta
decay with two electrons in the final state and nothing else, lepton number
violating decay modes can arise where new scalar 
\cite{Burgess:1992dt,Burgess:1993xh,Bamert:1994hb}
or even vector particles \cite{Carone:1993jv}
are emitted
as well:
\begin{equation}
2n \rightarrow  2p + 2 e^- + \phi,  2n \rightarrow  2p + 2 e^- + 2 \phi.  
\end{equation}
Majorons have been originally introduced as Nambu-Goldstone bosons
being responsible for breaking a global lepton symmetry and generating neutrino Majorana masses
\cite{Chikashige:1980ui,Gelmini:1980re}.
However, since such classical Majorons require severe fine-tuning in 
order to respect the bounds on neutrino masses and
at the same time induce an observable rate for neutrinoless double beta 
decay,  
several new models have been 
proposed in which the term Majoron refers more broadly to a light or 
massless boson with couplings to neutrinos.
In comparison to neutrinoless double beta decay 
with two electrons in the final state only, these decay modes 
lead to continuous spectra for the emitted electrons which can be discriminated from the Standard Model 
$2\nu\beta\beta$ decay by fitting the spectral shape. 
While expected decay rates are rather small due to suppressed
nuclear matrix elements 
\cite{Hirsch:1995in} experimental bounds 
from the Heidelberg-Moscow experiment using $^{76}$Ge
on various models have been derived in
\cite{Bockholt:1995gu}.
More recently,
the NEMO-3 
 \cite{Arnold:2006sd}
and KAMLAND-Zen \cite{Gando:2012pj}
collaborations have published 
improved bounds on these modes, constraining the 
 Majoron-neutrino coupling constant 
 from measurements in $^{100}$Mo, $^{82}$Se and $^{136}$Xe 
 e.g. for classical models
 to $\left\langle g_{ee} \right\rangle < (0.4-1.9) \cdot 10^{-4}$, $ \left\langle g_{ee} \right\rangle < (0.66-1.7) \cdot 10^{-4}$,
and 
$\left\langle g_{ee} \right\rangle < (0.8 - 1.6) \cdot 10^{-5}$, respectively,
depending on the value of the nuclear matrix element used.
For classical Majorons it has been shown in \cite{Tomas:2001dh} that by combining bounds from next-generation double beta experiments with constraints derived from supernova energy release arguments  
\cite{Kachelriess:2000qc} neutrino-Majoron couplings could be constrained down to a level of $10^{-7}$.

%------------------------------------------------------------------------------
\subsection{Non-Standard Neutrinos}
\label{sec:sterileneutrinos}
%------------------------------------------------------------------------------

Non-standard neutrinos can contribute to neutrinoless double beta decay
either as new flavours in addition to three partners of the
charged leptons included in the Standard Model, or via non-standard
properties or interactions. 

Additional flavours have to be either sterile or heavier than half the mass of th $Z$ boson in order to comply with the width of the $Z$ boson.
They will contribute to neutrinoless double beta decay via
mixing with the electron flavour neutrino (compare the discussion in
the contribution of Rodejohann). An update on a heavy 4th generation's
neutrino to neutrinoless double beta decay is given in \cite{Lenz:2011gd}.
It implies that such heavy active neutrinos have to be pseudo-Dirac
neutrinos with only a tiny amount of lepton number violation.
The amount of lepton number violation is further constrained due to wash-out
effects such weak scale pseudo-Dirac neutrinos will have on any 
pre-existing baryon asymmetry \cite{Hollenberg:2011kq}.

In principle the neutrino can also be a composite object. In this case a bound on the compositeness scale can be 
obtained from the neutrinoless double beta decay half life limit evaluated in the mass mechanism
\cite{Takasugi:1995bb,Takasugi:1997jd,Panella:1997wa}.

In
\cite{Uehara:2002wc} it has been discussed that observable 
double beta decay rates in extra-di\-mensional spacetimes could be
triggered by the lepton number violation induced by virtual black holes
violating the associated global symmetry.
A somewhat related mechanism arises in
theories with a saturated black hole bound on a large number of species.
Such theories
have been advocated recently as a possible solution to the 
hierarchy problem and as an explanation of the smallness of neutrino masses 
\cite{Dvali:2009ne}.
Then the violation of lepton number can create a potential phenomenological 
problem of such N-copy extensions of the Standard Model 
as again due to the lowered quantum gravity scale black holes may induce TeV scale 
LNV operators generating unacceptably large rates of e.g. neutrinoless double beta decay. 
It has been shown, however, that this does not happen in such a scenario due to a specific compensation mechanism between contributions of different Majorana neutrino states to these processes. As a result rates of LNV processes are extremely small and far beyond experimental reach, at least for the left-handed neutrino states
\cite{Kovalenko:2010ti}.

Lorentz invariance
and the equivalence principle are the most important 
pillars of special and general relativity. 
However, certain versions of string theories allow for or even predict the 
violation of these laws. Often a violation of Lorentz invariance (VLI)
implies that 
different particles can have characteristic maximal attainable 
velocities. The difference of the velocities $\delta v$ 
then parametrizes the size of VLI.
Similarly, the 
corresponding observable describing violations of the equivalence 
principle (VEP) is the difference of characteristic couplings 
$\delta g$ to the gravitational potential $\phi$. 
While previous studies of neutrino  oscillations provide
very restrictive bounds in 
the region of large mixing
\cite{Leung:1999cz}
$0\nu\beta\beta$ decay yields in certain models of VLI and VEP
a bound also in the 
region of zero mixing being not accessible to neutrino oscillation 
experiments\cite{KlapdorKleingrothaus:1998hm}: 
$\delta v<3.3 \cdot 10^{-16}$,
$\phi \cdot \delta g<3.3 \cdot 10^{-16}$.
Lorentz invariance is also closely related to CPT invariance.
In CPT violating models the conventional notion of Majorana neutrinos being 
their own anti-particles does not 
apply anymore. Majorana masses and double beta decay in such scenarios
are discussed in \cite{Barenboim:2002hx}.

%------------------------------------------------------------------------------
\section{Summary and Discussion}
\label{sec:conclusions}
%------------------------------------------------------------------------------

Neutrinoless double beta decay is a crucial observable in search for physics beyond the Standard Model as it tests the fundamental symmetry of lepton number. The violation of lepton number is predicted in many models of new physics and most prominently, it provides the only probe of the absolute mass scale of light Majorana neutrinos. In this context, searches for $0\nu\beta\beta$ are highly complementary to neutrino oscillation experiments, direct neutrino mass determinations in Tritium decay and cosmological observations of the impact of neutrinos on large scale structure formation. If the recent results at the LHC pointing to a Higgs boson with a mass of about 125~GeV are confirmed, we can expect to make a giant leap in understanding the nature of mass generation at the electroweak scale. Even then, the nature and the lightness of neutrinos would still be unexplained and so far neutrinoless double beta decay is the only realistic probe to distinguish between the Dirac or Majorana nature of light neutrinos. 

As outlined in Section~\ref{sec:EffectiveParametrization}, there is large number of possible effective operators that can give rise to $0\nu\beta\beta$ decay. Only a small selection of fundamental physics models that give rise to such LNV operators were presented in Section~\ref{sec:NewPhysicsModels}. Given these ambiguities, a crucial problem is to distinguish between different mechanisms. One possibility is to compare results from $0\nu\beta\beta$ with other neutrino experiments and cosmological observations. More generally, searches for physics beyond the Standard Model at other experiments, such as the LHC, can be correlated with $0\nu\beta\beta$ in specific models. More directly, it is also possible to infer the dominant mechanism purely within the context of neutrinoless double beta decay and closely related processes. Relevant techniques discussed in the literature include (i) the comparison of $0\nu\beta\beta$ decay rates in different isotopes \cite{Bilenky:2004um, Deppisch:2006hb, Gehman:2007qg} (this possibility is discussed in detail in the contribution by Fogli et al.), (ii) the comparison of $0\nu\beta^-\beta^-$ with $0\nu\beta^+\beta^+$ \cite{Hirsch:1994es}, (iii) the comparison of $0\nu\beta\beta$ with electron capture \cite{Hirsch:1994es}, (iv) the comparison of of $0\nu\beta\beta$ decay to the ground state and an excited state \cite{Simkovic:2001qf} and (v) the experimental determination of the angular and energy distribution of the outgoing electrons in $0\nu\beta\beta$ \cite{Doi:1982dn, Doi:1985dx, Tomoda:1986yz, Ali:2006iu, Ali:2007ec, Arnold:2010tu, Deppisch:2010zza}.

\begin{figure}[t]
\centering
\includegraphics[clip,width=0.40\textwidth]{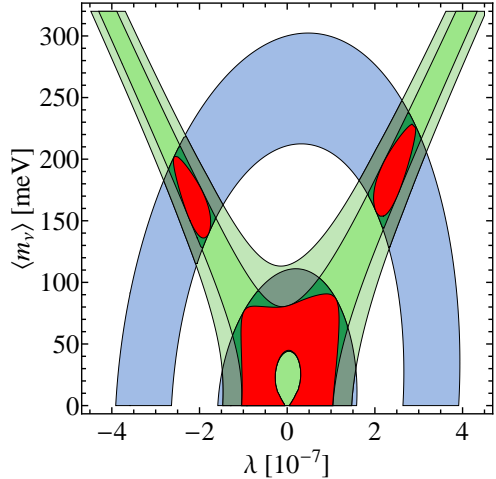}
\caption{Simulated constraints at one standard deviation on the effective neutrino mass $m_\nu$ and LRSM coupling $\langle\lambda\rangle \equiv \epsilon_{V+A}^{V+A}$ for the isotope $^{82}$Se at SuperNEMO from: (1) an observation of $0\nu\beta\beta$ decay half-life at $T_{1/2}=10^{25}$~y (outer blue elliptical contour) and $10^{26}$~y (inner blue elliptical contour); (2) reconstruction of the angular (outer, lighter green) and energy difference (inner, darker green) distribution shape; (3) combination of (1) and (2) (red contours). The strength of the $\lambda$ contribution is assumed to be 30\% of the standard mass mechanism. NME uncertainties are assumed to be 30\% and experimental statistical uncertainties are determined from the simulation in \cite{Arnold:2010tu}.}
\label{fig:discovery}
\end{figure}
Here we highlight the potential of measuring the angular and energy distribution in $0\nu\beta\beta$ tracking experiments such as SuperNEMO \cite{Arnold:2010tu}. Both the angular and energy correlation of the electrons depends on the effective operator mediating neutrinoless double beta decay. In the standard mass mechanism with $V-A$ couplings, the electrons are mostly emitted back to back with comparable energies. This is very different from the $\epsilon_{V+A}^{V+A}$ process shown in Fig.~\ref{fig:diagramsLR_LR}~(left), where the electrons are preferably emitted in the same direction with one electron taking away the majority of the energy. Fig.~\ref{fig:discovery} shows the impact of combining the total $0\nu\beta\beta$ rate measurement with a determination of the angular and energy distribution shape at SuperNEMO. In the hypothetical scenario considered, both the standard mass mechanism and the operator of $\epsilon_{V+A}^{V+A}$ are assumed to contribute; whereas the total rate alone can not differentiate between the two contributions, taking into account the angular and energy shape can allow to pinpoint their relative size.

If neutrinoless double beta decay is observed in next generation experiments at a level corresponding to an effective $0\nu\beta\beta$ mass $m_{ee} \gtrsim 10^{-2}$~eV, it would be an indication that the decay is caused by the exchange of light neutrinos, especially if corroborated by cosmological observations. On the other hand, such a conclusion is not straightforward as there is a large number of models which can trigger neutrinoless double beta decay. As discussed in this article, such new physics mechanisms can be economically categorized in terms of effective Lorentz invariant operators. Due to the black box theorem \cite{Schechter:1981bd, Nieves:1984sn, Takasugi:1984xr}, the observation of neutrinoless double beta decay will prove that the light left-handed neutrinos are Majorana particles, but from this alone it is not possible to infer the dominant operator and mechanism.

%------------------------------------------------------------------------------
\section*{Acknowledgements}
%------------------------------------------------------------------------------

M.H. acknowledges support from the Spanish MICINN grants FPA2011-22975, MULTIDARK CSD2009-00064 and by the Generalitat Valenciana grant Prometeo/2009/091 and the EU~Network grant UNILHC PITN-GA-2009-237920.

%%%%%%%%%%%%%%%%%%%%%%%%%%%%%%%%%%%%%%%%%%%%%%%%%%%%%%%%%%%%%%%%%%%%%%%%%%%%%%

\bibliographystyle{h-physrev4}
\bibliography{0vbb_focusitem_bsm}

\end{document}